\documentclass[a4paper,11pt]{article}
\usepackage{pos}

 \csname
@addtoreset\endcsname{equation}{section}

\def\mso{\mathfrak{so}}

\def\msp{\mathfrak{sp}}

\def\Real{{\mathbb R}}
\def\Comp{{\mathbb C}}

\def\bec{\begin{center}}
\def\ec{\end{center}}

\def\a{\alpha} \def\ad{\dot{\a}} 

\def\b{\beta}  \def\bd{\dot{\b}} 
\def\g{\gamma} 
\def\G{\Gamma}
\def\d{\delta} \def\dd{\dot{\d}}
\def\D{\Delta}
\def\e{\epsilon}

\def\vark{\varkappa}
\def\varkb{\bar{\varkappa}}
\def\l{\lambda}\def\tl{\tilde{\lambda}}
\def\L{\Lambda}
\def\m{\mu}
\def\n{\nu}

\def\s{\sigma}

\def\t{\tau}

\def\y{\eta}

\def\O{\Omega}
\def\o{\omega}

\def\kb{{\bar \kappa}}
\def\yb{{\bar y}}

\def\tx{\tilde{x}}

\def\ty{\widetilde{y}}
\def\tP{\widetilde{P}}

\def\tx{\widetilde{x}}

\def\tm{\widetilde{\m}}
\def\ty{\widetilde{y}}

\def\nn{\nonumber}
\newcommand{\eq}[1]{(\ref{#1})}

\def\be{\begin{equation}}
\def\ee{\end{equation}}
\def\bea{\begin{eqnarray}}
\def\eea{\end{eqnarray}}
\def\ba{\begin{array}}
\def\ea{\end{array}}

\def\seq{\;\;\equiv\;\;}

\def\ft#1#2{{\textstyle{{\scriptstyle #1}
\over {\scriptstyle #2}}}}

\def\scs#1{\section{\bf{\sc \large #1}}}
\def\scss#1{\subsection{\bf{\sc  #1}}}

\def\ad{\dot\alpha}
\def\bd{\dot\beta}

\def\au{{\underline{\alpha}}}
\def\bu{{\underline{\beta}}}

\newcommand\R{\mathbb{R}}
\newcommand\C{\mathbb{C}}

\title{Unfolding, higher spins, metaplectic groups\\ and resolution of classical singularities}
\ShortTitle{Unfolding and classical singularities}

\author*[a]{Carlo Iazeolla}
\author[b]{Per Sundell}

\affiliation[a]{Dipartimento di Scienze Ingegneristiche,\\ Guglielmo Marconi University -- Via Plinio 44, 00193, Roma, Italy \ \  \& \\
Sezione INFN Roma ``Tor Vergata'' -- Via della Ricerca Scientifica 1, 00133, Roma, Italy}

\affiliation[b]{Centro de Ciencias Exactas, Universidad del B\'io-B\'io,\\ 
Avda. Andr\'es Bello 720, 3800708, Chill\`an, Chile}

\emailAdd{c.iazeolla@gmail.com, c.iazeolla@unimarconi.it}
\emailAdd{per.sundell@unab.cl}

\abstract{We review and extend some recent results concerning the analysis of spacetime singularities in four-dimensional higher spin gravity, summarizing how the coupling of the gravitational field to massless higher spins may provide resolution mechanisms. We elucidate such mechanisms at the level of curvature singularities and degenerate metrics in exact as well as linearized solutions to Vasiliev's equations. As a preamble, we review the underlying higher-spin algebra and its metaplectic group extensions, after which we detail various gauge functions encoding the $AdS_4$ vacuum and the non-rotating Ba\~nados--Gomberoff--Martinez (BGM) metric, the four-dimensional lift of the spinless BTZ black hole, in different coordinate patches related by transition functions. We then revisit how, within the unfolded formalism, it is natural extend the BGM black hole through its causal singularity. Finally, we compare the metric-like and unfolded descriptions of scalar fluctuations over the (extended) BGM background, showing how the latter description maps singularities to well-defined metaplectic group elements providing regular values for the Weyl zero-form master field, which thus admits continuation
over the full extended BGM spacetime. }

\FullConference{%
  Corfu Summer Institute 2021 "School and Workshops on Elementary Particle Physics and Gravity"\\
  29 August - 9 October 2021\\
  Corfu, Greece
}

\tableofcontents

\begin{document}
\maketitle

\scs{Introduction}

One of the earliest and long-standing motivations to study higher-spin fields is to find out whether coupling the gravitational field to them may resolve classical spacetime singularities. 
This expectation found more concrete roots in the relatively simple example of supergravity, with the introduction of the gravitino, and then later in the UV properties of string theory, crucially involving an infinite tower of massive higher-spin fields. So a system of intermediate complexity like higher-spin gravity, describing the dynamics of an infinite multiplet of gauge fields of all spins --- that can be thought of as the first Regge trajectory collapsed to vanishing mass --- is a natural candidate theory in which to study this problem. 

Nonetheless, while strictly constrained by its infinite-dimensional local symmetry, higher-spin gravity is a very challenging theory to grasp, due to the fact that some degree of non-locality in the theory seems inescapable, and that the standard riemannian geometric setup, based on spin-$2$ constructs, has no invariant meaning and is to be replaced by a higher-spin extension thereof. However, precisely these properties make higher-spin gravity an especially interesting system in which to re-examine the status of spacetime singularities already at the classical level.

It is therefore especially fitting and instructive to attack this problem within the mathematical framework that has been built to handle the peculiarities of higher-spin physics in an efficient way: Vasiliev's non-linear system \cite{vasiliev,Vasiliev:1990vu,more,Vasiliev:2003ev,review99,Bekaert:2005vh,Iazeolla:2008bp,Didenko:2014dwa}, which encodes a highly complicated interacting gauge theory into a compact set of first-order differential constraints for a set of differential forms, referred to as \emph{master fields}, living on a fibered non-commutative extension of the spacetime manifold, sometimes referred to as \emph{correspondence space}. The evolution along the additional, non-commutative base directions generates the interaction vertices among physical fields, packed into the master fields together with auxiliary fields that absorb their derivatives. The formulation of the dynamics in terms of a Cartan-integrable set of zero-curvature and covariant constancy conditions, that the Vasiliev system is based on, is called \emph{unfolded formulation} \cite{Vasiliev:1988sa,Vasiliev:1992gr,Vasiliev:2012vf,Tarusov:2021fdk} and can be thought of as a covariant analogue of hamiltonian dynamics. While  it may superficially look inconvenient with respect to the standard framework of non-abelian gauge theories, unfolding has a number of  powerful consequences, that in fact enabled the formulation of higher-spin gravity in closed form: two of the most important ones are that the gauge invariance of the vertices is a consequence of the integrability of their generating system in correspondence space; and that including the interactions as solutions of a differential constraints in auxiliary variables $Z$, with gauge and field-redefinition ambiguities encoded into the choice of resolution operator for the $Z$-dependence, gives some mathematical tool to control the resulting spacetime non-locality of the vertices and, possibly, to come up with a generalization of that concept adapted to higher-spin gravity (see \cite{Vasiliev:2017cae,Didenko:2018fgx,Didenko:2019xzz,Gelfond:2019tac,Didenko:2020bxd,Gelfond:2021two,COMST,meta} for recent progresses). 

But unfolding is a formulation available for any dynamical system and provides powerful methods to address many other physical and mathematical questions (see, e.g., among many others, \cite{Shaynkman,Horava:2007ds,fibre,Skvortsov:2008vs,BMV1,BMV2,Boulanger:2011dd,Boulanger:2014vya,Boulanger:2015mka}). Not only it makes the gauge symmetries of the problem manifest, with all differential forms appearing in unfolded equations by construction filling modules of the symmetry algebra; it can also incorporate gravity without singling out the metric nor requiring its inverse. Moreover, once the individual forms are packed into the master fields --- depending on spacetime and fibre coordinates, $x$ and $Y$, respectively --- subject to zero-curvature and covariant constancy conditions, to a large extent the spacetime features of the solutions become stored in their dependence on fibre coordinates --- in a sort of spacetime/fibre duality much akin to a Penrose transform \cite{Vasiliev:2012vf}. At the linearized level, this translates into a clean separation of the building blocks of solutions, corresponding to the moduli:  fibre representatives of the Weyl zero-form master field, carrying the local degrees of freedom of the solutions; gauge functions entirely absorbing the spacetime dependence, responsible for possible boundary degrees of freedom; holonomies of the vacuum connection; and windings in the transition functions gluing master fields over different charts. These features make unfolding a potentially very efficient tool for exploring the systematics of solution spaces (see \cite{Corfu} and references therein), and to address the problem of spacetime singularities. 

In this paper we shall review and extend some recent results concerning and interlacing these issues. After recalling in Sections \ref{sec:HSalg} and \ref{sec:unfolding} some aspects of higher-spin algebras, metaplectic group and unfolding that will be of relevance for our analysis, we set the stage for tackling spacetime singularities, in particular curvature singularities and degenerate metrics. We thus devote Section \ref{sec:vacua} to a somewhat detailed study of gauge functions encoding vacuum solutions to the four-dimensional bosonic Vasiliev equations and a few relevant transition functions, going beyond the results so far appeared in the literature. We begin by giving the gauge functions for the $AdS_4$ background in different coordinate systems. In order to exhibit all the ingredients of the unfolding formulation at work in a simple example, we also provide the transition functions gluing two stereographic charts, and show some of their peculiarities: in particular, we write an improper Lorentz transformation (a hyperplane reflection) by means of (holomorphic) metaplectic group elements.  

We then move on to the four-dimensional analogue of the non-rotating BTZ black hole, first constructed by Aminneborg, Bengtsson, Holst and Peldan \cite{Aminneborg:1996iz}, and then later revisited by Ba\~nados, Gomberoff and Martinez (BGM) \cite{BGM}, who properly interpreted it as a constantly curved black hole with geometry ${\rm CMink}_3\times_\xi S^1$ (where ${\rm CMink}$ denotes a conformal Minkowski spacetime, and $\times_\xi$ a warped product) that traps circles.  

%(add here observation on transition functions?)

In the standard, metric-like description, black holes of this type are obtained from identifying points in $AdS$, along a non-compact Killing vector $\vec K$ with $\xi^2:=\vec K^2$, from an ambient-space construction, a procedure which leads naturally to cutting off a portion of spacetime, leaving $\xi\geq 0$ and leading to geodesic incompleteness and a degenerate frame field at $\xi=0$. As we shall see, in the unfolded formalism, working intrinsecally at the level of gauge functions, it will be natural to extend the BGM manifold through the singularity ($\xi\gtreqless 0$, still excluding closed timelike curves), reaching an extended BGM black hole which can be described as the gluing of two non-rotating BGM black holes along their past
and future space-like singularities \cite{ourBTZ}. This extension can in fact be described, for a certain choice of coordinates, as originating from an analytic continuation in the gauge function. 

Then, in Section \ref{sec:fluct} we compare the metric-like and the unfolded description of fluctuations over the BGM spacetime. In particular, we recall how the unfolded formalism permits the construction of fluctuation fields from fibre representatives, defined in coordinate-free bases, and we show with an explicit example in what sense the singular behaviour of a scalar field at the BGM singularity is encoded and resolved at the level of the fluctuation master field, which remains well-defined as the frame field degenerates and hence admits continuation across singularities and over the full extended BGM spacetime \cite{ourBTZ}.  

In order to show the latter result, we make use of an observation originally made \cite{2011} in the study of 4D higher-spin spherically-symmetric black holes \cite{Didenko:2009td,2011,2017}. These are solutions of the full (as well as linearized, at least in certain generalized gauges) theory comprising a tower of AdS Schwarzschild-like Weyl tensors of all spins, each Weyl tensor of spin $s$ carrying a $r^{-s-1}$ dependence and thus blowing up at the origin. However, as we shall briefly review, the ill behaviour of the individual spin-$s$ Weyl tensors translates to a delta-function behaviour of the corresponding master field at the singularity, and distributions in non-commutative variables can be considered smooth since they have good star product properties. Indeed, delta functions of non-commutative variables are equivalent to bounded functions up to a change in the ordering prescription \cite{meta}. It is in this same sense that the Weyl zero-form master field over the BGM background remains well-defined even where individual fluctuation fields are irregular. 

The paper is completed by two appendices, in which we collect our spinor and $AdS$ conventions and identify and characterize a few relevant elements of the metaplectic group.

Summarizing, the results reported and extended in this paper provide examples of how the possible  higher-spin resolution of classical spacetime singularities relies not only on the higher-spin  extension of gravity, but crucially on its implementation using Vasiliev’s unfolded formulation in terms of master fields, with the spacetime/fibre duality that it entails. Indeed, it is only by working intrinsically, with field equations formulated as a differential graded algebra and gauge functions and fibre representatives of solutions as main building blocks, that we are able to envisage singularity-resolution mechanisms that are unattainable in the ordinary, metric-like formalism, and that seem to reduce certain type of singularities to artifacts of the basis choice of the fibre operator algebra.

\scs{Higher-spin algebra and metaplectic groups}\label{sec:HSalg} 

\paragraph{Higher-spin algebra.} The basic building block of higher-spin gravities in four spacetime dimensions with  negative cosmological constant is Dirac's conformal particle on the real hypercone with signature $(2,3)$.
Its quantization provides a left-module $|{\cal S})$ for the associative higher-spin algebra  \cite{Fradkin:1986ka,Vasiliev:1986qx,Konstein:1989ij} 
\be {\cal H}:= \frac{{\rm Env}[\mso(2,3)]}{{\rm Anni}[|{\cal S})]}\ ,\ee
formed by quotienting the unital universal enveloping algebra of the Lie algebra $\mathfrak{so}(2,3)$ by the annihilator of $|{\cal S})$, which is the ideal in ${\rm Env}[\mso(2,3)]$ generated by 
\be V_{AB}:=\frac12 M_{(A}{}^C\star M_{B)C}-\frac1{10} \eta_{AB}C_2\approx 0\ ,\qquad V_{ABCD}:=M_{[AB}\star M_{CD]}\approx 0\ ,\label{1.2}\ee
with $C_2:=\frac12 M^{AB}\star M_{AB}$, where $\star$ denotes the associative product, and $M_{AB}=-M_{BA}$, $A\in\{0',0,1,2,3\}$, are $\mathfrak{so}(2,3)$ generators obeying $(M_{AB})^\dagger=M_{AB}$ and 
\begin{align}
  \left[M_{AB},\,M_{CD}\right]_\star
  &=
  i\left(\eta_{AD}M_{BC}+\eta_{BC}M_{AD}-\eta_{AC}M_{BD}-\eta_{BD}M_{AC}\right)
  \ ,
\end{align}
with $\eta_{AB}={\rm diag}(-,-,+,+,+)$.

\paragraph{Adjoint and twisted-adjoint representations.} 
The higher-spin algebra acts on itself through twisted-adjoint actions
\be {\rm ad}_{\alpha,\beta}(P_1) P_2:= \alpha(P_1) \star P_2-P_2\star\beta(P_1)\ ,\qquad P_1, P_2\in {\cal H}\ ,\ee
where $\alpha$ and $\beta$ are  $\mso(2,3)$-morphisms; as these act faithfully on any subspace of ${\rm Env}[\mso(2,3)]$ preserved under the adjoint $\mso(2,3)$-action, including ${\rm Anni}[\cal S]$, they lift to morphisms of ${\cal H}$.
These actions induce ${\cal H}$-modules
\be {\cal T}_{\alpha,\beta}:= ({\cal H},{\rm ad}_{\alpha,\beta})\ , \qquad [{\rm ad}_{\alpha,\beta}(P_1),{\rm ad}_{\alpha,\beta}(P_2)]={\rm ad}_{\alpha,\beta}([P_1,P_2]_\star)\ .\ee
The adjoint module ${\cal T}:={\cal T}_{{\rm Id},{\rm Id}}$ has a decomposition 
\be {\cal T}\downarrow_{{\rm ad}(\mathfrak{so}(2,3))}=\bigoplus_{n=0}^{\infty}{\cal T}^{[n,n]}\ ,\ee
into irreducible $\mso(2,3)$-tensors ${\cal T}^{[n,n]}$ consisting of monomials in $M_{AB}$ of degree $n$ projected onto the Young tableaux of highest weight $(n,n)$.
Defining transvections $P_{a}:= M_{0'a}$ obeying
\be [P_a,P_b]_\star=i M_{ab}\ ,\ee
where $M_{ab}$ generate the Lorentz $\mso(1,3)\subset \mso(2,3)$ stabilized by the automorphism $\pi$ defined by
\begin{align}
  \pi(P_a):=-P_a\ ,
\end{align}
the corresponding twisted-adjoint module $\widetilde{\cal T}\equiv {\cal T}_{{\rm Id},\pi}$ has a decomposition
\be \widetilde{\cal T}\downarrow_{\widetilde{\rm ad}(\mso(2,3))}=\bigoplus_{s=0}^\infty \widetilde{\cal T}^{[s,s]}\ ,\ee
into infinite-dimensional $\mso(2,3)$-irreps $\widetilde{\cal T}^{[s,s]}$ with further decomposition 
\be \widetilde{\cal T}^{[s,s]}\downarrow_{\widetilde{\rm ad}(\mso(1,3))}\equiv \widetilde{\cal T}^{[s,s]}\downarrow_{{\rm ad}(\mso(1,3))}=\bigoplus_{k=0}^\infty \widetilde{\cal T}^{[s,s;s+k,s]}\ ,\ee
into irreducible Lorentz tensors $\widetilde{\cal T}^{[s,s;s+k,s]}$ built from $s$ powers of ${M}_{ab}$ and $k$ powers of $P_a$ projected onto the Young tableaux with highest $\mso(1,3)$-weight $(s+k,s)$.
Remarkably, the Casimir operator $C_2$ obeys \cite{fibre}
\be C_2[{\cal T}^{[n,n]}]=C_2[\widetilde{\cal T}^{[n+1,+1]}]\ ,\qquad n=0,1,\dots\ .\ee

\paragraph{Oscillator realization.} 
From Eq. \eqref{1.2} it follows that
\be C_2[\mathfrak{so}(2,3)||{\cal S})]\approx -\frac{5}4\ ,\ee
coinciding with its value in the oscillator representation of the Lie algebra $\mathfrak{sp}(4;\mathbb{C})\cong\mathfrak{so}(5;\C)$.
This representation arises naturally in the holomorphic symplectic $\C^{4}$, viewed as a differential Poisson manifold with trivial pre-connection \cite{Arias:2015wha}, which deforms the unital differential graded associative algebra of holomorphic polynomial forms on $\mathbb{C}^{4}$ into a non-commutative ditto with product $\star$.
Letting $(Y^{\underline\alpha},\overline{Y}^{\bar{\underline{\alpha}}})$, with $\underline{\alpha},\bar{\underline{\alpha}}=1,\dots,4$, be complex canonical coordinates in which the two-form is given by $C+C^{\dagger_\C}$, where $C:=\frac12 dY^{\underline\alpha} \wedge dY^{\underline{\beta}} C_{\underline{\alpha\beta}}$ the  hermitian conjugation operation reads
\be (Y^{\underline\alpha})^{\dagger_\C}=\overline Y^{\bar{\underline\alpha}}\ ,\qquad d\circ \dagger_\C=\dagger_\C\circ d\ ,\ee
and the graded non-commutative holomorphic algebra is generated by $(Y^{\underline{\alpha}},dY^{\underline\alpha})$ modulo
\be \left[Y^{\underline\alpha},\,Y^{\underline\beta}\right]_\star=2iC^{\underline{\alpha\beta}}\ ,\qquad \left[Y^{\underline\alpha},dY^{\underline\beta}\right]_\star=0\ ,\qquad \left[dY^{\underline\alpha},dY^{\underline\beta}\right]_\star=0\ ,\ee
where $C^{\underline{\alpha\beta}}C_{\underline{\alpha\gamma}}=\delta^{\underline{\beta}}_{\underline{\gamma}}$.
Denoting its degree-zero subalgebra, referred to as the holomorphic Weyl algebra, by ${\cal P}[\mathbb{C}^{4}]$, and letting $\Gamma_{\cal H}\cong \mathbb{Z}_2\times \mathbb{Z}_2$ be the discrete subgroup of ${\rm Diff}(\C^4)$ generated by the involutive automorphisms $\Pi,\gamma: \C^4\to \C^4$ of the holomorphic differential Poisson structure defined by
\be Y^{\underline\alpha}\circ \Pi:= -Y^{\underline\alpha}\ ,\quad Y^{\underline\alpha}\circ \gamma:= \Gamma^{\underline{\alpha}}_{\bar{\underline{\alpha}}}\,\overline Y^{\bar{\underline\alpha}}\ ,\qquad \Pi\circ \gamma=\gamma\circ \Pi\ ,\ee
one identifies the higher-spin algebra as\footnote{If a group $G$ acts on a space $V$, then $V^G$ denotes the set of elements in $V$ that are invariant under $G$.}
\be {\cal H}\cong  {\cal P}[\R^4]:=\left({\cal P}[\mathbb{C}^{4}]\otimes ({\cal P}[\mathbb{C}^{4}])^{\dagger_\C}\right)^{\Gamma_{\cal H}}\ ,\ee
that is, the Weyl algebra of complex polynomials on the noncommutative $\R^4$ obtained by deforming the differential Kaehler structure with two-form $\frac12 (C+C^\dagger)$, using the hermitian conjugation operation
\be (Y^{\underline\alpha})^{\dagger}=\Gamma^{\underline{\alpha}}_{\underline{\beta}}\, Y^{{\underline\beta}}\ ,\qquad d\circ \dagger=\dagger\circ d\ ;\ee
in particular, letting $(\Gamma_A)_{\underline{\alpha}}{}^{\underline{\beta}}$ be Dirac matrices of $\mso(2,3)$ obeying $\Gamma_A \Gamma_B=\eta_{ab}+\Gamma_{AB}$, one has
\be M_{AB}=\frac18 Y \Gamma_{AB}  Y\ ,\ee
using conventions in which $U^{\underline{\alpha}}:=C^{\underline{\alpha\beta}}U_{\underline{\beta}}$ and $UM V:= U^{\underline{\alpha}}M_{\underline{\alpha}}{}^{\underline{\beta}}V_{\underline{\beta}}$ (see Appendix \ref{app:conv} for our spinor and $\mso(2,3)$ conventions).

\paragraph{Holomorphic metaplectic group.} Strict quantization of ${\cal P}[\C^4]$ in left-modules $|{\cal V})$ equipped with non-degenerate $\mathfrak{sp}(4;\C)$-invariant bilinear forms, making $|{\cal V})\cong ({\cal V}^\ast|$ yield operator algebras ${\rm End}(|{\cal V}))$ whose elements can be sent by Wigner--Ville maps to classical distributions on spaces of test functions on families of planes $\R^{4}\subset\C^{4}$, referred to as symbols, forming associative algebras 
\be {\cal A}_{\cal V}[\C^{4}]\cong {\rm End}(|{\cal V}))\ ,\ee
with composition rules defined by letting the symbols act on themselves via twisted convolution formulae given by integrals over the $\R^{4}\subset\C^{4}$; for example, in the Weyl ordering scheme
\be
  ({f_1}\star {f_2})(Y)=
  \int\frac{d^{4}Y_1\,d^{4}Y_2}{(2\pi)^{4}}
  \,e^{iY_1Y_2}
  f_1(Y+Y_1)f_2(Y-Y_2)
  \ ,
\ee
for $f_1, f_2\in {\cal A}_{\cal V}[\C^{4}|\R^{4}]$. 
Viewed as an infinite-dimensional manifold, ${\cal P}[\mathbb{C}^4]$ admits a complex structure compatible with the star-product, which can be extended\footnote{The extension is non-trivial since if $\Omega'\subset \R^{N'}$ is non-compact and $f:\Omega\times \Omega'\to \C$ is analytic on $\Omega\subset \C^{N}$, then the integral $\int_{v\in \Omega'} d^{N'} v f(u,v)$ need not depend analytically on $u$.} to ${\cal A}[\C^{4}|\mathbb{R}^{4}]$ by compactifying the auxiliary integrals, yielding the complex metaplectic double cover \cite{meta}
\be \mathbb{Z}_2 \rightarrow Mp(4;\C)\stackrel{\rm Pr}{\longrightarrow} Sp(4;\C)\ee
of $Sp(4;\C)$, with holomorphic projection map
\begin{align} R(g)\star Y^{\underline{\alpha}}\star R(g)^{\star(-1)}&:= Y^{\underline{\beta}} ({\ Pr}(g))_{\underline{\beta}}{}^{\underline{\alpha}}\ ,\end{align}
where $R: Mp(4;\mathbb{C})\to {\cal A}[\C^{4}|\R]$ is the holomorphic representation map determined by the analytical continuation of
$U: Sp(4;\C)|_{\rm cut}\to {\cal A}[\C^4]$ given by
\be
U(S)
  :=
\frac{ 1}{\sqrt{\det\frac{1+S}{2}}}
  \exp\left(
  \tfrac{i}{2}Y\tfrac{1-S}{1+S}Y
  \right)
  \ ,\label{1.11}
\ee 
using the coordinatization of $Sp(4;\C)$ in terms of $S\in {\rm mat}_{4}(\C)$ obeying
\begin{align}
  S_{\underline\alpha}{}^{\underline\alpha'}\;S_{\underline\beta}{}^{\underline\beta'}\;C_{\underline{\alpha'\beta'}}
  &=
  C_{\underline{\alpha\beta}}\ ,
\end{align}
which furnishes a projective representation, viz.
\be
  U(S_1) \star U(S_2)
  =
  e^{i\varphi(S_1,S_2)}
  U(S_1 S_2)
  \ ,
\ee
with cocycle $\varphi: Sp(4;\C)|_{\rm cut}\times Sp(4;\C)|_{\rm cut}\to \{0,\pi\}$ obeying 
\be \varphi(S_1,S_2)+\varphi(S_1 S_2,S_3)-\varphi(S_1,S_2S_3)-\varphi(S_2,S_3)=0\ .\ee
Thus, as a manifold,
\be Mp(4;\C)= Mp(4;\C)_+\cup Mp(4;\C)_-\ ,\qquad Mp(4;\C)_\pm \stackrel{\rm top}{\cong} Sp(4;\C)|_{\rm cut}\ ,\ee
with
\be R(g_\pm)
  =\pm U({\rm Pr}(g_\pm))
  \,,\qquad g_\pm\in Mp(4;\C)_\pm\ .\ee
The projective representation can be constructed by first composing $\exp: \mathfrak{sp}(4;\C)\to Sp(4;\C)$ with the oscillator realization  
\be  \label{eq:repr sp}
 M(\Theta):=
  -\frac{i}{4}Y \Theta Y
  \,,\qquad
  \left[M(\Theta_1),\,M(\Theta_2)\right]_\star
  =
  M\left([\Theta_1,\,\Theta_2]\right)
  \,,
\ee
of $\mathfrak{sp}(4;\C)\cong {\rm Sym}_{4}(\C)$, yielding\footnote{$\exp_\star A$ denotes a star-power expansion $\exp_\star A=1+A+\ft12 A\star A+...$ .}. 
\be U(e^{-2\Theta})=\exp_\star \left(-2M(\Theta)\right) =
  \frac{1}{\sqrt{\det \cosh \Theta}}
  \exp\left(
  \tfrac{i}{2}Y \tanh \Theta Y\right)
  \ ,\label{1.17}
\ee
where the pre-factor is defined using the branch cut; while $\exp(\mathfrak{sp}(4;\C))$ is a proper subset of $Sp(4;\C)$, Eq. \eqref{1.17} can be continued analytically to Eq. \eqref{1.11}, and then further to $R$, which is thus defined independently of the choice of branch cut \cite{meta}.

\paragraph{Real metaplectic subgroup.} According to Bargmann's theorem, which states that a unitary representation of a Lie group $G$ with trivial $\pi_1(G)$ can be de-projectivized, it follows from $\pi_1(Sp(4;\C))=\{e\}$ that any unitary representation of $Sp(4;\C)$ is non-projective; for example, $S\mapsto U(S)\star ({U}(S^{-1}))^{\dagger_\C}$ provides a unitary non-projective representation of $Sp(4;\C)$ in $|{\cal V})\otimes |\overline{\cal V})$ with a realization in terms of symbols in ${\cal A}_{\cal V}[\C^4]\otimes ({\cal A}_{\cal V}[\C^4])^{\dagger_\C}$. 
Conversely, from $\pi_1(Sp(4;\R))=\mathbb{Z}$ it follows that the restriction of $R$ to the real metaplectic group 
\be \mathbb{Z}_2\rightarrow Mp(4;\R)\stackrel{Pr}{\longrightarrow} Sp(4;\R)\ ,\ee
defined by $Mp(4;\R)=\{g\in Mp(4;\C)|{\rm Pr}(g)\in Sp(4;\Real)\}$, yields a unitary irreducible $\mathbb{Z}_2$-projective metaplectic, or Segal--Shale--Weil, representation \cite{Folland,Guillemin:1990ew,Woit:2017vqo} of $Sp(4;\mathbb{R})$ in $|{\cal V})$, equipped with an $\mathfrak{sp}(4;\C)$-invariant, positive definite, sesquilinear form.
Thus, $g\in Mp(4;\R)$ is realized by a symbol 
\be R(g)\in {\cal A}_{\cal V}[\R^4]:=\left({\cal A}_{\cal V}[\C^4]\otimes ({\cal A}_{\cal V}[\C^4])^{\dagger_\C}\right)^{\Gamma_{\cal H}}\ ,\ee
obeying $(R(g))^\dagger=(R(g))^{\star(-1)}$; indeed, the restriction of $U$ to the topological $S^1\subset Sp(4;\R)$ is double-valued.

\paragraph{Inner Klein operators.} 
It follows from Eq. \eqref{1.11}, that limits of $R(g)\in {\cal A}[\C^{4}|\R^{4}]$ in which $1+{\rm Pr}(g)$ degenerates are analytic delta sequences \cite{meta}; for details, see Appendix \ref{App:proof}.
In particular, the center
\be Z(Mp(4;\C))=\{I_\pm,K_\pm\}\ ,\ee
whose elements obey
\be (I_-)^2=(K_+)^2=(K_-)^2=I_+={\rm Id}\ ,\qquad K_+ K_-=I_-\ ,\ee
with metaplectic representation 
\be
 R(I_\pm|Y)=\pm 1
  \ ,\qquad R(K_\pm|Y)=\pm K_Y
  \ ,\qquad  K_Y:=(2\pi)^2\delta^{4}(Y)\ ,\ee
obeying 
\be {\rm Pr}(I_\pm)=I_{4\times 4}\ ,\qquad {\rm Pr}(K_\pm)=-I_{4\times 4}\ ,\ee
from which it follows that $\Pi\in \Gamma_{\cal H}$ has an inner realization in ${\cal A}_{\cal V}[\R^4]$, viz.
\be K_Y\star Y^{\underline\alpha}= - Y^{\underline\alpha}\star K_Y\ ,\qquad K_Y\star K_Y=1
  \ .
\ee
The Kaehler structure on $\R^4$ is equivalent to a holomorphic symplectic structure on $\C^2$ that can be exhibited by splitting
\be Y^{\underline{\alpha}}=(y^\alpha,\bar y^{\dot{\alpha}})\ ,\qquad (y^\alpha)^\dagger=\bar y^{\dot{\alpha}}\ ,\ee
and defining the complex metaplectic subgroup $Mp(2;\C)\subset Mp(4;\R)$ by
\be {\rm Pr}(g)_{\underline{\alpha}}{}^{\underline\beta}=\left[\begin{array}{cc} {\rm Pr}(g)_{{\alpha}}{}^{\beta}&0\\0& \delta_{\dot\alpha}{}^{\dot\beta}\end{array}\right]\ ,\qquad g\in Mp(2;\C)\ ,\ee
with center
\be Z(Mp(2;C))=\{i_\pm,k_\pm\}\ ,\label{2.40}\ee
obeying
\be (i_-)^2={\rm Id}\ ,\qquad (k_+)^2=(k_-)^2=i_-\ ,\qquad i_- k_+=k_-\ ,\qquad k_+ k_-=i_+\ ,\ee
\be
 R(i_\pm|y)=\pm 1\, \qquad R(k_\pm)=\mp i\kappa_y\ ,\qquad \kappa_y=2\pi \delta^2(y)\ ,\ee
idem ${\overline Mp}(2;\C)={\rm Stab}_{Mp(4;\R)}(Mp(2;\C))$ and $\{\bar i_\pm,\bar k_\pm\}$, from which it follows that the automorphism $\pi$ has an inner realization in ${\cal A}_{\cal V}[\R^4]$
as well, viz.
\be \kappa_y\star y^{\alpha}= - y^{\alpha}\star \kappa_y\ ,\qquad \kappa_y\star \kappa_y=1
\ ,
\ee
and that the Klein operator $K_Y$ can be factorized holomorphically in ${\cal A}_{\cal V}[\R^4]$, viz.
\be K_Y=\kappa_y\star \bar\kappa_{\bar y} \ ,\ee
as a consequence of $K_\pm= k_\pm \bar k_\pm$.

\paragraph{Projectors at infinity.} As a manifold, the metaplectic group $Mp(4;\C)$ can be extended to a compact space ${Mp}_\infty(4;\C)$ by adding points $p_\infty$ at infinities such that 
\be \lim_{g\to p_\infty} R(p|Y)=0\ ,\ee
corresponding to projectors 
\be P(p_\infty|Y):= \lim_{g\to p_\infty} {\cal N}(g) R(g|Y)\ ,\qquad P(p_\infty|Y)\star P(p_\infty|Y)=P(p_\infty|Y)\ ,\label{1.43}\ee
where ${\cal N}:Mp(4;\C)\to \C$ diverge at $p_\infty$ so as to cancel the evanescent prefactor in $U({\rm Pr}(g)|Y)$, leaving a uniquely determined normalization constant.
Thus, defining ${\cal N}':Sp(4;\C)\to \C$ by ${\cal N}'({\rm Pr}(g))={\cal N}(g)$ for $g\in Mp(4;\C)$, one can define a compactification ${Sp}_\infty(4;\C)$ of $Sp(4;\C)$ such that
\be \lim_{S\to S_\infty} {\cal N}'(S) U(S|Y):= P(p_\infty|Y) \ ,\ee
and view the set of projector points as the ramification points of the holomorphic projection map.
The massless-particle and black-hole states in global $AdS_4$ arise from ${\cal H}$-orbits of such projector points, as we shall recall in Section \ref{sec:fluct}.

\scs{Unfolded formulation} \label{sec:unfolding} 

By introducing frame fields and sufficiently many auxiliary fields, any set of partial differential equations can be formulated as a Cartan integrable system (CIS) of zero-curvature conditions on a set of of differential forms \cite{Vasiliev:1988sa,Vasiliev:1992gr,review99,Vasiliev:2012vf,Tarusov:2021fdk} forming a locally defined free differential algebra (FDA) \cite{Sullivan,vN,DAuria:1982mkx}. 
Conversely, the original equations resurface in regions with non-degenerate frame, where the auxiliary fields can be decomposed into Lorentz tensors that can be either expressed in terms of derivatives of the original dynamical fields, or set to zero by fixing local shift symmetries.

The resulting approach to dynamical systems, referred to as unfolded dynamics, is manifestly diffeomorphism invariant, which facilitates the study of field theory in regions where the metric, hence causal structure, degenerates.
Moreover, as local degrees of freedom arise as integration constants of infinite-dimensional towers of zero-forms related to covariant Taylor expansions of matter fields and on-shell curvatures, unfolded dynamics can be used to map singularities to states in infinite-dimensional representations of the gauge algebra, which paves the way for resolving these types of singularities in the higher-spin context.
Indeed, the unfolded formulation of higher-spin gravity is manifestly gauge invariant, given in terms of form fields in various linear representations of the higher-spin algebra, which can be expanded dual  bases adapted to the nature of the Weyl curvature.

\paragraph{Local formulation.} Restricted to a chart $U$ of a manifold, an unfolded system is described by a set $\{W^A\}$ of locally defined differential forms generating a FDA, that is, they obey a CIS of generalized curvature constraints
\be R^A:=dW^A+Q^A(W) \ = \ 0 \ ,\label{CIS} \ee
where $Q^A$ are exterior polynomials in the form fields obeying structure equations\footnote{A CIS on a chart $U$ can alternatively be viewed as the equation of motion for a  Alexandrov--Kontsevich--Schwarz--Zaboronsky sigma model \cite{AKSZ} in which the forms on $U$ are mapped to functions on $T[1]U$ given on-shell by pull-backs of coordinates on a graded target space equipped with a vector field $\vec Q := Q^A \vec\partial_A$ in degree one that is nilpotent, viz. $\vec Q^2  \equiv  0 $.}
\be Q^B\wedge \frac{\partial Q^A}{\partial W^B} \ \equiv \ 0 \ ,\ee
independently of the dimension of $U$, which ensure the generalized Bianchi identities
\be dR^A- R^B\wedge \frac{\partial Q^A}{\partial W^B}\equiv 0\ .\ee
It follows that Eq. \eqref{CIS} is
not only compatible with $d^2\equiv 0$, but also explicitly integrable on $U$ by applying finite Cartan gauge transformations to locally defined zero-form integration constants.
Letting $p_A$ denote the degree of $W^A$, the linearized Cartan gauge transformations
\be \d_\e W^A \ = \ T^A_\e:=d\e^A-\e^B\wedge \frac{\partial Q^A}{\partial W^B} \ , \label{Cartangt} \ee
where $\e^A$ are gauge parameters of degree $p^A-1$, induce linear transformations of the Cartan curvatures, viz.
\be \delta_\e R^A\equiv -(-1)^{p_B} R^B\wedge \e^C \wedge \partial_C\partial_B Q^A\ .\ee
It follows that if $C^A=\delta_{0,p_A} C^A$ are constants, then
\be W^A_{\lambda;C}:=\left. \left[\exp(T^B_\lambda\partial_B) W^A \right]\right|_{W=C}\ ,\label{CIS2}\ee
solve Eq. \eqref{CIS}, and conversely any locally defined classical solution must be of the form \eqref{CIS2}.

\paragraph{Linearized higher-spin gravity.}

Vasiliev's equations describe FDAs on noncommutative manifolds generated by locally defined dynamical forms in degrees zero and one, and a globally defined closed and central two-form\footnote{The system is a consistent truncation of a flat superconnection comprising dynamical forms of degrees zero, one and two \cite{FCS}.}.
These algebras can be reduced\footnote{The reduction requires boundary conditions on the connection along the noncommutative directions; for details, see \cite{COMST,meta}.} to subalgebras defined locally on charts of a commutative manifold $M$ in terms of a set of perturbatively defined unfolded Fronsdal fields.
The reduced systems can be further expanded around locally constantly curved gravitational backgrounds with coordinate-free descriptions in terms of one-form connections $\Omega\in \msp(4;\Real)$ obeying
\be d\O+\O\star \O  \ = \ 0\ ,\qquad \Omega^\dagger=-\Omega\ . \label{flat}\ee
Focusing on the model with higher-spin algebra ${\cal H}$, the unfolded description of its linearized fluctuations around $\Omega$ requires a twisted-adjoint zero-form $\Phi \in \widetilde{\cal T}$, referred to as the Weyl zero-form, and an adjoint one-form $W\in {\cal T}$, obeying 
\be\label{COMST} D^{(0)}\Phi=0\ ,\qquad
D^{(0)} W+\Sigma(e,e;\Phi)=0\ , \ee
where the covariant derivatives
\be D^{(0)}\Phi:=d\Phi+\Omega\star \Phi-\Phi\star\pi(\Omega)\ ,\qquad D^{(0)} W:= dW+\Omega\star W+W\star \Omega\ ,\ee 
and the twisted-adjoint zero-form module is glued to the adjoint one-form module via the cocycle 
\be \Sigma(e,e;\Phi):=
\frac{ib}{4}\left.e^{\a\ad}\wedge e_{\a}{}^{\bd}\partial^\yb_{\ad} \partial^\yb_{\bd} \Phi\right|_{y=0}
+\frac{i\bar b}{4}\left.e^{\a\ad}\wedge e^{\b}{}_{\ad}\partial^y_\a \partial^y_\b \Phi\right|_{\bar{y}=0}
\,,\ee
using a decomposition 
\be \O=e+\o\ ,\qquad e:=-i e^a P_a\ ,\qquad \omega:=  -\frac{i}2\o^{ab}M_{ab}\ ,\label{AdSO}\ee
of $\Omega$ into a Lorentz connection $\omega$ and transvection gauge field $e$, which thus obey $\pi(\omega)=\omega$ and $\pi(e)=-e$, and $b,\bar b$ are phases that can be fixed by requiring parity invariance \cite{Sezgin:2003pt}.
Eqs. \eq{flat} and \eq{COMST} form a CIS, with abelian gauge symmetries associated to $W$, leaving $\Omega$ and $\Phi$ inert, and nonabelian gauge symmetries associated to $\Omega$, under which $\Phi$ and $W$ transform in twisted-adjoint and adjoint representations, respectively.
Finally, higher-spin Killing symmetries arise as background gauge symmetries leaving $\Omega$ inert. 
Imposing reality conditions
\be \Phi^\dagger=\pi(\Phi)\ ,\qquad W^\dagger=-W\ ,\label{reality}\ee
defining Lorentz-covariant derivatives 
\be \nabla \Phi=d\Phi+[\o,\Phi]_\star\ ,\qquad \nabla W=dW+[\o,W]_\star\ ,\ee
and decomposing 
\be \Phi= \sum_{s\geqslant 0} \Phi^{[s,s]}\ ,\qquad W=\sum_{s\geqslant 1} W^{[s,s]}\ ,\ee
where $\Phi^{[s,s]}\in \widetilde{\cal T}^{[s,s]}$ and $W^{[s,s]}\in {\cal T}^{[s,s]}$, Eq. \eqref{COMST} decomposes into unfolded equations of motion for a real scalar field, viz.
\be\nabla\Phi^{[0,0]}+e\star \Phi^{[0,0]}+\Phi^{[0,0]}\star e=0\ ,\ee
and a tower of real Fronsdal fields of ranks $s=1,2,3,...$, viz.
\bea \nabla W^{[s,s]}+e\star W^{[s,s]}+W^{[s,s]}\star e + \Sigma^{[s,s]}(e,e;C^{(s,s)})&=&0\ ,\\ \nabla\Phi^{[s,s]}+e\star \Phi^{[s,s]}+\Phi^{[s,s]}\star e&=&0\ ,\eea
with cocycles
\be \Sigma^{[s,s]}(e,e;\Phi^{(s,s)})=
\frac{ib}{4}e^{\a\ad}\wedge e_{\a}{}^{\bd}\,\Phi_{\dot\alpha \dot\beta\dot\gamma(2s-2)} \bar y^{\dot\gamma(2s-2)}
+{\rm h.c}
\,.\ee
In a region $U$ where $e$ defines a non-degenerate Lorentz frame, the gauge fields can be converted into irreducible Lorentz tensors, and the constraints into algebraic equations for auxiliary fields and second-order differential equations.
As a result, set of the component fields that are algebraically independent modulo curvature constraints and local shift symmetries, consists of the scalar field
\be C:= \Phi|_{Y=0}\ ,\ee
and the Fronsdal fields
\be C_{\alpha(s)\dot\a(s)}:= (e^{-1})_{\alpha\dot\alpha}{}^\mu \left.\frac{\partial^{2s}}{\partial y^{\alpha(s)}\partial \bar y^{\dot\alpha(s)}} W_\mu\right|_{Y=0}\ ,\qquad s=1,2,\dots\ ,\ee
where $\mu$ denotes a world index on $U$.
Among the auxuliary fields are 
\be \Phi_{\a(2s)}=\left.\frac{\partial^{2s}}{\partial y^{\alpha(2s)}} \Phi\right|_{Y=0}\ ,\qquad s=1,2,\dots\ , \label{Phicomp}\ee
and their hermitian conjugates, making up the selfdual and anti-selfdual components of the Faraday tensor $\Phi_{a,b}$ for $s=1$, the linearized Weyl tensor $\Phi_{ab,cd}$ for $s=2$, and higher-spin generalized linearized Weyl tensors $\Phi_{a(s),b(s)}$ for $s\geqslant 3$, where $a(n):=(a_1...a_n)$ and the Weyl tensors are traceless for $s\geqslant 2$.
The corresponding Klein--Gordon, Maxwell and Bargmann--Wigner equations for $s\geqslant 2$ read \cite{review99,Bekaert:2005vh,Didenko:2014dwa,Iazeolla:2008bp,fibre,BMV1}
\bea
  s=0&\ :\quad&
  (\nabla^2 +2)C
  = 0\ , \label{2.30}\\
  s=1&\ :\quad &\nabla^a \Phi_{a,b}
  = 0
  \ ,\qquad 
  \nabla_{[a} \Phi_{b,c]}
 = 0\ ,\\
 s\geqslant 2&\ :\quad&
  \nabla_{[a} \Phi_{b|b_2\dots b_s,|c]c_2\dots c_s}
  = 0\ ,
\label{2.32}
\eea
which are thus equivalent to $D^{(0)}\Phi=0$ iff $e$ is non-degenerate.

\paragraph{Local spacetime/fibre duality.}
Without any non-degeneracy assumption on $e$, Eqs. \eqref{flat} and \eqref{COMST} can be solved by introducing gauge functions \cite{Vasiliev:1990bu}
\be L_\xi: M_\xi \to R({Mp}(4;\R)/Z({Mp}(4;\R)))\ ,\ee
and twisted-adjoint integration constants
\be \Phi'_\xi\in {\cal A}_{\cal V}[\R^4]\ ,\qquad d\Phi'_\xi=0\ ,\ee
both of which are defined locally on charts $M_\xi$ of $M$, such that
\be
\Omega_\xi|_{M_\xi} \ = \ L_\xi^{\star(-1)}\star dL_\xi
\ ,\qquad \Phi_\xi|_{M_\xi} \ = \ L_\xi^{\star(-1)}\star \Phi'_\xi\star \pi(L_\xi)\ .\label{LrotPhi}\ee
Introducing the adjoint initial data
\be \Psi':=\Phi'\star\kappa_y\ ,\ee
in terms of which
\be \Phi_\xi|_{M_\xi}= L_\xi^{\star(-1)}\star \Psi'_\xi\star L_\xi\star \kappa_y\ ,\ee
makes it manifest that the locally defined solutions are invariant under redefinitions 
\be L_\xi \sim R(Z_\xi) \star L_\xi\ ,\qquad Z_\xi\in Z(Mp(4;\R))\ .\ee
We refer to the locally defined solution as a \emph{regular} unfolded field configuration if
\be \Phi_\xi|_{M_\xi}\in \widetilde{\cal T}\ ,\label{regular}\ee
and $\Omega_\xi$ is bounded.
To construct regular configurations, one may start by assuming the existence of an \emph{unfolding point} $p_\xi\in M_\xi$ where \cite{review99,Bekaert:2005vh}
\be L_\xi|_{p_\xi}=1\ ,\qquad\Phi_\xi|_{p_\xi}=\Phi'_\xi\ ,\ee
after which $L_\xi$ can be deformed homotopically in the interior of $M_\xi$ so as to impose the regularity condition on $\Phi_\xi$ and $\Omega_\xi$, which amounts to resolving a locally defined singularity if ${\cal A}_{\cal V}[\R^4]\cap \widetilde{\cal T}=0$.

In a region where $e_\xi$ is non-degenerate, Eq. \eqref{LrotPhi} 
thus maps the local degrees of freedom of the linearized theory, that is, all local information that is invariant under abelian gauge transformations, to the operator algebra ${\cal A}_{\cal V}[\R^4]$.
Conversely, the gauge function $L_\xi$ spreads, or unfolds, the local datum $\Psi'_\xi$, which we hence refer to as the \emph{initial datum}, or \emph{fibre representative} of the linearized solution, over the spacetime chart $M_\xi$.

\paragraph{Killing parameters.} A higher-spin Killing symmetry parameter $\e^{(0)}$ obeys
\be D^{(0)} \e^{(0)}=0\ ,\qquad \e^{(0)}\in {\cal T}\ . \label{Keq}\ee
Using gauge functions, it follows that 
\be \e^{(0)}=L^{\star(-1)}\star \e^{\prime(0)}\star L\,,\qquad d\e^{\prime(0)}=0\ ,\qquad \e^{\prime (0)}\in {\cal T}\ , \label{Lkill}\ee
suppressing chart indices.
The adjoint action of $L$ on a symbol $f(Y)$ amounts to a rotation of the oscillators, viz.
\be f^L(Y)~:=~ L^{\star(-1)}\star f(Y)\star L \ = \ f(Y^L)\ ,\ee\be  Y^{L}_{\underline\a}~:=~L^{\star(-1)}\star Y_{\underline\a}\star L~=~ L_{\underline\a}{}^{\underline\b} Y_{\underline\b}\ ,\label{YLrotn}\ee
where $L_{\underline{\a\b}}\in Sp(4;\R)$.
In particular, the spin-two Killing parameter corresponding $M_{AB}$ is given by
\be  M_{AB}^L \ = \ \frac18 Y\Gamma_{AB}^L Y \ , \qquad  (\Gamma_{AB}^L)_{\underline{\a\b}} \ = \ -(L^T\Gamma_{AB} L)_{\au\bu} = \ \left(\ba{cc} \vark^L_{\a\b} & v^L_{\a\bd} \\[5pt] \bar v^L_{\ad \b} & \bar \vark^L_{\ad\bd}\ea\right)_{AB} \ ;\label{Gammakv}\ee
if $e$ is non-degenerate, then the off-diagonal blocks yield Killing vector fields $\vec v^L= v^L_{\a\dot\a} (e^{-1})^{\a\dot\a,\mu}\vec \partial_\mu$, and the diagonal ones yield the (anti-)selfdual components of the corresponding Killing two-form (see e.g. \cite{Didenko:2009tc}, and \cite{Corfu} for a few examples). 

\paragraph{Global formulation.}
Globally defined solutions are constructed by selecting a structure subgroup 
\be H\subset Mp(4;\Real)/Z(Mp(4;\R))\ ,\ee
and patching together the locally defined configurations via
\be L_\xi= C_\xi^{\xi'}\star L_{\xi'}\star T_{\xi'}^{\xi}\ ,\qquad \Phi_\xi \ = \ T_\xi^{\xi'}\star \Phi_{\xi'}\star \pi(T_{\xi'}^{\xi}) \ ,\label{phitrans}\ee
using transition functions
\be T_\xi^{\xi'}: M_\xi\cap M_{\xi'}\to R(H)\ ,\qquad T_\xi^{\xi}=1\ ,\ee
obeying triple-overlap conditions $T_\xi^{\xi'}\star T_{\xi'}^{\xi''}\star T_{\xi''}^{\xi}=1$, and gauge-function integration constants acting on the zero-form integration constants, viz.
\be \Psi'_\xi=C_\xi^{\xi'}\star \Psi'_{\xi'}\star C_{\xi'}^{\xi}\ ,\qquad dC_\xi^{\xi'}=0\ , C_\xi^{\xi'}\in R(Mp(4;\R)/Z(Mp(4;\R))\ .\ee
It follows that by redefinitions of the transition functions one may take
\be L_\xi:M_\xi \to R(Mp(4;\R)/H)\ .\ee
The choice of $H$ influences the abundance of classical observables of the theory \cite{2005,Sezgin:2011hq,2011}, as these functionals must be manifestly invariant under local gauge transformations with parameters from $H$.
Of particular interest are holonomies 
\be {\rm Hol}:\gamma\in \pi_1(M)\mapsto {\rm Hol}_\gamma(\Omega)=P \exp_\star \oint_\gamma \Omega\ ;\ee
cutting into open portions $\gamma_i\in M_{\xi(i)}$ such that $\gamma=\gamma_1 \#\gamma_2\cdots\# \gamma_N$, one has
\be  {\rm Hol}_\gamma(\Omega)\equiv P\prod_{i=1}^{N} T_{\xi(i+1)}^{\xi(i)}\star \exp_\star \oint_{\gamma_i} \Omega_{\xi(i)}=\prod_{i=1}^{N} C_{\xi(i+1)}^{\xi(i)}\ \ee
A natural choice of structure group is the metaplectic extension of the Lorentz group defined by
\be H:=\left\{t\in Mp(4;\Real)/Z(Mp(4;\R))\big|k_+ t k_+= t\right\}\ ,\ee
where $k_+$ is defined below Eq. \eqref{2.40}, which we shall use in what follows; it follows that if $t\in H$, then
\be R(t)^\dagger= R(t)^{\star(-1)}\star R(z)\ ,\qquad \pi(R(t))=R(t)\star R(z')\ ,\qquad z,z'\in Z(Mp(4;\R))\ .\ee
In summary, the linearized solution spaces are built from the following moduli\footnote{Interactions generally complicate the picture: for instance, implementing specific boundary conditions in the full Vasiliev system requires simultaneous, field-dependent adjustements of gauge function and Weyl zero-form initial data \cite{COMST,Corfu}.}:
\begin{enumerate}
\item zero-form integration constants encoding local degrees of freedom;
\item boundary values of gauge functions encoding boundary degrees of freedom; 
\item gauge function integration constants encoded into holonomies;
\item windings in transition functions encoded into structure group Chern classes.
\end{enumerate}
In particular, given a fixed vacuum configuration, the boundary conditions on the fluctuations are thus mapped to algebraic properties of the zero-form integration constants, and it is in this sense that a spacetime/fibre duality \cite{Vasiliev:2012vf,Corfu} is established by unfolding.

\vspace{0.5cm}
We shall next turn to exhibiting the building blocks above described and the spacetime/fibre duality in concrete cases where holonomies are activated, that also show how unfolding provides a powerful tool to deal with certain types of singularities.

\scs{Higher-spin gravity vacua}\label{sec:vacua} 

In this section, we describe classical solutions of Vasiliev's equations on four-manifolds of various topologies with non-trivial first homotopy groups, in which the Weyl zero-form vanishes and the higher-spin connection supports non-trivial holonomies.

\scss{Global $AdS_4$ spacetime}

The global $AdS_4$ spacetime has topology $M\cong S^1\times \R^3$ and non-degenerate frame field, corresponding to the metric induced by embedding the spacetime into a flat five-dimensional spacetime with metric $ds^2  =dX^A dX^B\eta_{AB}$ as the hyperboloid 
\be X^A X^B \eta_{AB}=-X_{0'}^2-X_0^2+X_1^2+X_2^2+X_3^2= -1 \label{hyperb}\ee 
assuming unit radius.
In what follows, we shall provide gauge functions for the corresponding $\mso(2,3)$-valued connection $\Omega$ corresponding to different choices of coordinate systems related to one another by transition functions.

\paragraph{Stereographic coordinates.} A convenient set of intrinsic coordinates are the stereographic coordinates $x^a_{\pm}$, $ a = 0,1,2,3 $, that cover $AdS$ by means of two charts $U_{\pm}$, and are related to the embedding coordinates $X^A$ via 
\bea   x^a_{\pm} \ = \  \frac{X^{a}}{1+|X^{0^{\prime }}|} \ , \qquad \left.|X^{0^{\prime
}}|\right|_{U_{\pm}}\ = \ \pm X^{0^{\prime }} \ , \label{stereo-embed1} \eea
with inverses
\bea (X^a,X^{0'})|_{U_\pm} \ = \ \left(\frac{2x^a_\pm}{1- x_\pm^2},\pm\frac{1+x^2_\pm}{1-x^2_\pm}\right)\ , \qquad -1<x^2_\pm <1 \ ,\quad x^2=x^a x^b\eta_{ab} \ . \label{stereo-embed2}\eea
The metric in stereographic coordinates takes a manifestly Lorentz-invariant form,
\be ds^2 \ = \ \frac{4dx^2}{(1-x^2)^2} \ , \qquad x^2\neq 1 \ ,\label{AdSstereo}\ee
which is left invariant by the inversion $x^a_\pm = -x^a_\mp/(x_\mp)^2$ that relates the two sets of stereographic coordinates in the overlap region $ (x_+)^2, (x_-)^2  <  0$. 
Inversion maps the future and past time-like cones into themselves and exchanges the two space-like regions $0< x^2< 1$ and $x^2 > 1$ while leaving the boundary $ x^2 =1$ fixed. %It follows that the single cover of $AdS_4$ is formally covered by taking $x^a\in \Real^{1,3}$.
 
The corresponding gauge function on each chart is 
\be L_{{\rm stereo}\pm} \ = \ \exp_\star(i\m^a(x_\pm)P_a) \ , \label{Lstereo} \ee
where 
\be \m^a(x_\pm) = 4\left({\rm arctanh}\sqrt{\frac{1-h_\pm}{1+h_\pm}}\right)\frac{x_\pm^a}{\sqrt{x^2_\pm}} \ ,  \qquad h_\pm:=\sqrt{1-x_\pm^2} \ , \label{mu1}\ee
with the equivalent useful rewriting
\be  {\rm arctanh}\left(\sqrt{\frac{1-h_\pm}{1+h_\pm}}\right) \ = \ \frac14 \ln\left(\frac{1+\sqrt{x^2_\pm}}{1-\sqrt{x^2_\pm}}\right) \ .  \label{mu2}\ee
That \eq{Lstereo} generates \eq{AdSstereo} is easily shown by using the Baker-Campbell-Hausdorff formula (in infinitesimal form): defining 
\be \l^a(x) \ := \ i\m^a \ = \ \l(x) \, n^a \ , \ee
where 
\be \l(x) \ = \ 4\,i\,{\rm arctanh}\sqrt{\frac{1-h}{1+h}} \ = \ i\,\ln\frac{1+\sqrt{x^2}}{1-\sqrt{x^2}} \ , \ee
and 
\be n^a \ = \ \frac{x^a}{\sqrt{x^2}} \ , \qquad n^a n_a \ = \ 1  \, \ee
one can write,
\bea L^{\star( -1)}\star dL & = & \frac{\sin\l}{\l}\,d\l^a P_a-\frac{\l\cdot d\l}{\l^2}\left(\frac{\sin\l}{\l}-1\right)\l^a P_a+i\,\frac{\cos\l-1}{\l^2}\l^a d\l^b M_{ab} \nn\\
&=& \sin\l \,dn^a P_a+d\l\, n^aP_a+i(\cos\l-1) n^a dn^bM_{ab} \ , \label{BCHfin}\eea
which, comparing with \eq{AdSO}, gives 
\be e^a  \ = \ -\frac{2dx^a}{1-x^2}\ , \qquad \o^{ab} \ = \ \frac{4x^{[a}dx^{b]}}{1-x^2} \ . \ee

The transition function mapping between the two charts and giving the right action under coordinate inversion on the vielbein,
\bea  e^a\qquad \xrightarrow[x^a\to -x^a/x^2 ] \ \ \ \ \   \L^{ab}e_b \ , \qquad  \L^{ab} \ = \ \eta^{ab}-2n^a n^b &    \label{improperL}\eea
can be written as
\bea T_{-+} \ = \ \kappa_y\star {\cal T}_+  \ , \qquad {\cal T}_+ \ := \ \exp_\star(-\pi n_+^a P_a) \ , \label{Tmp}\eea
where $n_\pm^a:=\frac{x^a_\pm}{\sqrt{x^2_\pm}}$, which indeed for any vector $v^a$ (with fibre indices) gives %
\be (T_{-+})^{\star( -1)}\star v^a P_a \star T_{-+} \ = \ \L^{ab}v_b P_a \ , \ee
as can be checked using $\kappa_y\star P_a=-P_a\star\kappa_y$ and the Baker-Campbell-Hausdorff formula for a generic $g=\exp_\star(q^aP_a)$, 
\bea g^{\star( -1)}\star v^a P_a \star g = (\cos q)\, v^aP_a+\frac{1-\cos q}{q^2} (q\cdot v) \,q^aP_a + i\frac{\sin q}{q}\,q^av^bM_{ab} \eea
applied to $g=\exp_\star(-\pi n^a P_a)$. $\L^{ab}$ is of course a Lorentz transformation performing reflection in the hyperplane orthogonal to $n^a$, well-defined for $x^2\neq 0$ and involutory. As for the action of $T_{-+}$ on $L$, note that the action of a transition function on the gauge function is, generally, defined up to a spacetime-constant involution $C$, e.g., $L_+=C\star L_-\star T_{-+}$, as the presence of such $C$ does not alter the action of $T_{-+}$ on the connection $L^{-1}\star dL$. Choosing $C=\kappa_y$ correctly gives\footnote{It is important to note that, while an involutory constant element $C$ is immaterial for the action of the transition function on the vacuum connection $L^{\star( -1)}\star dL$, it acts non-trivially on the Weyl zero-form initial data: indeed, denoting $L_{{\rm stereo}\pm}$ with $L_{\pm}$, from \eq{phitrans} referred for simplicity to the adjoint initial datum $\Psi'=\Phi'\star\kappa_y$, we can see that in order for $\Psi_+=(T_{-+})^{\star( -1)}\star \Psi_-\star T_{-+}$ to hold together with $\Psi_+=L_+^{\star( -1)}\star\Psi_+\star L_+$ and $\Psi_-=L_-^{\star( -1)}\star\Psi_-\star L_-$, then the realization \eq{TonL} implies $\Psi'_+=\kappa_y\star\Psi'_-\star\kappa_y$. This gives one concrete simple example of the non-trivial interplay of initial data and gauge functions in realizing a given solution in different charts.}
\be L_{{\rm stereo}+} \ = \ \kappa_y\star L_{{\rm stereo}-}\star T_{-+} \ . \label{TonL}\ee
Indeed, considering that $n_-^a=-n_+^a$, and using ${\rm arg}(-1)=\pi$ and the definition \eq{Tmp}, 
\bea \kappa_y\star L_{{\rm stereo}-}\star T_{-+} & = &  \exp_\star i\left[\ln\left(\frac{1+\sqrt{x_+^2}}{1-\sqrt{x_+^2}}\right)-i\pi\right] n_+^a P_a \ \star \  \exp_\star(-\pi n^a_+ P_a) \nn\\ & = & \exp_\star i\left(\ln\frac{1+\sqrt{x_+^2}}{1-\sqrt{x_+^2}}\right)n_+^a P_a \ = \ L_{{\rm stereo}+} \ .\eea
From \eq{TonL} it is easy to see that the action of the transition function on the gauge function becomes trivial at $x^2_+=x^2_-=-1$, where indeed $x_+^a=x_-^a$: indeed, $\left.L_{{\rm stereo}-}\right|_{x^2_+=x^2_-=-1}=\left. L_{{\rm stereo}+}\right|_{x^2_+=x^2_-=-1}=\exp_\star (i\frac{\pi}2 x_+^a P_a)$, which is indeed left invariant under \eq{TonL}. At the same time, $\left. T_{-+}\right|_{x^2_+=x^2_-=-1}\neq 1$, which is a manifestation of the fact that this gauge function is an improper Lorentz transformation, not connected to the identity, coherently with the property $\det \L=-1$ of the reflection matrix \eq{improperL}.

The transition function $T_{+-}$ can be defined analogously, with exchange of the roles of $\pm$, and up to an element of the centre $\pm\kappa_y\star\kb_{\yb}$. This means that one can equivalently take
\bea &  T_{+-} \ = \ \pm\kb_{\yb}\star {\cal T}_-  \ , \qquad {\cal T}_- \ := \ \exp_\star(-\pi n_-^a P_a) \ , &\nn\\ 
& {\rm and} \qquad L_{{\rm stereo}-}  =  \mp\kappa_y\star L_{{\rm stereo}+}\star T_{+-}\ , & \label{Tpm1} \eea
or
\bea  & T_{+-} \ = \ \pm\kappa_y\star {\cal T}_-  \ ,& \nn\\
& {\rm and} \qquad L_{{\rm stereo}-}  =  \mp\kb_{\yb}\star L_{{\rm stereo}+}\star T_{+-} \ ,&
\label{Tpm2} \eea
 as in both cases nesting twice the transformations for $L_{{\rm stereo}\pm}$ gives the identity. This can be easily checked by making use of the identity
\be {\cal T}_\pm \star {\cal T}_\pm \ = \ \exp_\star(-2\pi n_\pm^a P_a) \ = \ -\kappa_y\star\kb_{\yb} \label{T-2pi} \ee
(see Appendix \ref{App:proof}). 

Recalling that $\kappa_y$ (which can be written as a star-exponential $\kappa_y=-i\exp_\star(i\pi w_y)$, where $w_y=\frac{i}{4}yRy$, with $R$ denoting a matrix such that $R^2=1$ \cite{meta}) is an element of $Mp(2,\C)$, this example shows how improper Lorentz transformations can be encoded into the product of an element of $Mp(2,\C)$ with elements of $Mp(4;\R)$ at specific, discrete points. Moreover, the reality properties of this product are actually those of an element of $Mp(4;\R)$ up to an element of the centre, $(T_{-+})^\dagger=-T^{\star( -1)}_{-+}$.

\vspace{0.3cm}

Note however that the gauge function can also be trivially extended to arbitrary $x_\pm^2<-1$, since $\m^a(x)$ is real for any $x^a\in \R^{1,3}$ such that $x^2<1$. Indeed, from its definition \eq{mu1}-\eq{mu2} it is clear that for any $x^2<0$, i.e. $\sqrt{x^2}\in i\R$, $\mu^a$ reduces to $ \m^a(x)=4x^a\,{\rm arctan}(\sqrt{\frac{h-1}{h+1}})/|\sqrt{x^2}|\in \R$. 

But the gauge function above can in fact be analytically continued even to $x^2>1$. In fact, while $\m^a(x)$ at the exponent becomes complex, due to the presence of $h=\sqrt{1-x^2}$ (or $\ln(1-x^2)$ in \eq{mu2}), 
\be \left.\l\right|_{x^2>1} \ = \ 4\,i\,{\rm arctanh}\sqrt{\frac{1-i\tilde{h}}{1+i\tilde{h}}} \ , \qquad \tilde{h}:= \sqrt{x^2-1} \ ,\ee
this has no consequence for the connection\footnote{As $\m^a(x)$ acquires an imaginary part for $x^2>1$, it may seem puzzling that the connection $L^{\star( -1)}\star dL$ remains antihermitian, as is the case for $x^2<1$, where $L^\dagger=L^{\star( -1)}$. One way to clarify this issue is to use \eq{mu2} to rewrite 
\be \left.L\right|_{x^2>1}= \exp_\star\left(i\ln\frac{\sqrt{x^2}+1}{\sqrt{x^2}-1}n^aP_a\right) \star \exp_\star(\pi n^a P_a) =:\widetilde{L}\star {\cal T}\ ,\nn\ee
and then observe that, as a consequence of \eq{BCHfin} applied to ${\cal T}$, 
${\cal T}^{\star( -1)}\star d{\cal T}$ is $\pi$-even, i.e., only has components on $M_{ab}$, 
$ {\cal T}^{\star( -1)}\star d{\cal T}  =  -2in^a dn^b M_{ab}$, 
which in particular implies that ${\cal T}\star d{\cal T}^{\star( -1)}={\cal T}^{\star( -1)}\star d{\cal T}$, and hence 
\bea \left((\left.L\right|_{x^2>1})^{\star(-1)}\star d(\left.L\right|_{x^2>1})\right)^\dagger &=& \left(\widetilde{L}^{-1}\star d\widetilde{L}+\widetilde{L}^{\star(-1)}\star {\cal T}^{-1}\star d{\cal T}\star \widetilde{L}\right)^{\dagger}=-\widetilde{L}^{\star( -1)}\star d\widetilde{L}-\widetilde{L}^{\star( -1)}\star {\cal T}\star d{\cal T}^{\star( -1)}\star\widetilde{L}\nn\\
&=& - \widetilde{L}^{\star( -1)}\star d\widetilde{L}-\widetilde{L}^{\star(-1)}\star {\cal T}^{-1}\star d{\cal T}\star\widetilde{L} = -(\left.L\right|_{x^2>1})^{\star( -1)}\star d(\left.L\right|_{x^2>1}) \ .\nn\eea
} $\O=L^{\star(-1)}\star dL$, as only integer powers of $1-x^2$ appear in \eq{BCHfin}.
%
%it is easy to see that, while 
%, as, e.g.,
%
%\be \sin \l \ = \ i \sinh\left(4\,{\rm arctanh}\sqrt{\frac{1-i\tilde{h}}{1+i\tilde{h}}}\right) \ = \ -2i\frac{\sqrt{1+\tilde{h}^2}}{\tilde{h}^2} \ = \ 2i\frac{\sqrt{1-h^2}}{h^2} \ = \ 2i\frac{\sqrt{x^2}}{1-x^2}\ .\ee
%
Thus, one can cover the entire $AdS_4$ with a single gauge function, analytically continued where $x^2>1$,
\be L_{\rm stereo} (x) \ = \ \exp_\star(i\m^a(x)P_a) \ , \qquad x^2 \neq 1 \ . \label{veryextLstereo}\ee

\paragraph{Spherical coordinates.} The familiar global spherical coordinates $(t,r,\theta,\varphi)$ in which the metric reads
\bea  ds^2 \ = \ -(1+r^2)dt^2+\frac{dr^2}{1+ r^2}+r^2(d\theta^2+\sin^2\theta d\varphi^2) \ ,\label{metricglob}\eea
are related locally to the embedding coordinates by
\bea & X^0 \ = \ \sqrt{1+r^2}\sin t \ , \qquad X^{0'} \ = \ \sqrt{1+r^2}\cos t \ , & \nn\\[5pt]
& X^1 \ = \ r\sin\theta\cos\varphi \ , \quad  X^2 \ = \ r\sin\theta\sin\varphi \ , \quad X^3 \ = \ r\cos\theta \ ,& \eea
providing a one-to-one map if $t\in [0,2\pi)$, $r\in[0,\infty)$, $\theta\in[0,\pi]$ and $\varphi\in[0,2\pi)$ defining the single cover of $AdS_4$. 

The gauge function for global $AdS_4$ in spherical coordinates $(t,r,n^i)$, $i=1,2,3$,  $n^i n_i=1$ is
\be L_{\rm spherical}=\exp_\star (iEt)\star \exp_\star (i \,n^i P_i \,{\rm arcsinh} r)\ ,\label{Lspher}\ee
where $E$ is the energy operator and $P_i$ are the spatial transvections in $\mathfrak{so}(2,3)$. The factorization of the $t$-dependence reflects the topology $S^1\times \mathbb{R}^3$ of global $AdS_4$, where $S^1$ is the closed timelike circle, and 
the periodicity in $t$ of the global $AdS_4$ connection\footnote{To avoid closed timelike curves it is customary to decompactify the time circle and work on the universal covering space of $AdS_4$, with topology $\R^4$.} is concretely manifested by the fact that its holonomy along $S^1$ is
\be  {\rm Hol}_{S^1}(\Omega) \ = \ \exp_\star(2\pi i E) \ = \ -\kappa_y\star \kb_{\yb} \ , \label{holkappa}\ee
which is a central element in $R(Mp(4;\R))$ (for a proof of the second equality in \eq{holkappa}, see Appendix \ref{App:proof}).

\scss{Spinless BGM black hole}

As is well known, pure 3D Einstein gravity can be thought of as a topological theory with structure group $SO(1,2)$ and dynamical field given by a on-shell flat one-form $\Omega$ valued in the Lie algebra $\mathfrak{g}$ of $G=SO(1,3)$, $SO(2,2)$ or $ISO(1,2)$ depending on whether the cosmological constant is positive, negative or null.
Despite the fact that any vacuum solution with a negative cosmological constant is locally $AdS_3$, letting go of the global invertibility of the vielbein one can construct topologically and  causally non-trivial solutions such as the celebrated BTZ black hole \cite{BTZ,BHTZ}. 

The same construction can be repeated any higher dimension \cite{Aminneborg:1996iz,Aminneborg:1997pz,BGM,Breview}, and in particular in the 4D case which we are interested in, giving rise to a class of \emph{constant curvature black holes}. 
All such spacetimes have in common the property that, while being locally trivial, they are geodesically incomplete, due to their peculiar topology of type $S^1\times \R^{D-1}$, in $D$ spacetime dimensions, where the $S^1$ is along a non-compact direction. 
While the higher-dimensional lift of the well-studied 3D BTZ black hole preserves such features rather straightforwardly, the lifting of the corresponding classical observables is problematic, as gravitational gauge fields in $D\geq 4$ are deformed on-shell by Weyl tensors which appear to obstruct any intrinsically defined functional that reduces on-shell to a holonomy \cite{Guilleminot:2017hrp,Guilleminot:2017hfy}. 

Vasiliev's higher-spin gravity, on the other hand, contains a flat one-form valued in a higher-spin algebra, even in the presence of a non-trivial Weyl zero-form.
Thus, the theory maps closed curves in spacetime to holonomies valued in the metaplectic group \cite{ourBTZ,COMST}, which can be used to characterize these BHTZ-like geometries upon embedding them into higher-spin gravity where they are naturally interpreted as topologically non-trivial vacua (see also \cite{LebBTZ} for the embedding of the BTZ black hole in 3D higher-spin gravity).

\paragraph{Ambient metric construction.} A natural way of constructing constantly curved black holes is the one first employed in the 3D BTZ case and then extended to higher dimensions: the non-trivial topology is induced via a quotient $AdS_D/\G$ of $AdS_D$ obtained by identifying points along the orbit of a non-compact Killing vector field $\vec K$, where $\Gamma\cong \mathbb{Z}$ is the discrete subgroup of the diffeomorphism group generated by $\exp{2\pi \overrightarrow K}$, corresponding to a group element  $\gamma\in SO(2,D-1)$. We shall from now on fix our attention on $D=4$. 

The non-rotating 4D BGM black hole arises from choosing, $\vec K$ to be one of the $AdS$ transvections, viz.
\be \vec K \ = \ \sqrt{M}\,\vec v_P \ , \label{vecK}\ee
where we denote with $\vec v_P $ the Killing vector associated with a transvection generator $P$, which we can for definiteness choose to be $P=P_{1}=M_{0'1}$, and $M$ is the mass of the BGM black hole; at the level of constructing a constantly curved black hole, one may equivalently consider a boost, though this degeneracy is broken at the level of fluctuations, as we shall exhibit below.  

To understand the geometric consequences of the identification along the orbits of $\vec K$ it is useful to refer to the embedding picture \eq{hyperb}, in which the Killing vectors are represented as $\vec {v}_{AB} = X_A \overrightarrow{\partial}_B - X_B \overrightarrow\partial_A$. As the norm of \eq{vecK} in the whole $AdS_4$ spacetime is indefinite, 
\be \xi^2 \equiv \overrightarrow K^2= M\left( (X^{0'})^2 - (X^1)^2\right) \ , \ee
the identification produces closed time-like curves in the region in which $\xi^2< 0$. It is thus natural to remove this region from the quotient spacetime $AdS_4/\G$, and in that sense the surface $\xi^2=0$, i.e. the two-sheeted hyperboloid
\be \xi^2= 0 \qquad \longleftrightarrow\qquad  X_0^2-X_2^2-X_3^2=1 \label{xi2=0}\ee
becomes a causal singularity, because geodesics terminate there; and the cone 
\be \xi^2=M \qquad \longleftrightarrow\qquad  X_0^2-X_2^2-X_3^2=0 \ , \ee
which \eq{xi2=0} asymptotes to, represents a horizon, as all future-directed geodesics from the points of its $X^0>0$ ($X^0<0$) branch can only hit (come from) the future (past) singularity \cite{BGM}. 

The resulting manifold can be naturally parameterized via an intrinsic coordinate system that is adapted to the Killing vector along which the identification is performed. Thus, introducing a coordinate $\phi$ such that $\vec {v}_{AB}=\vec \partial_\phi$, a natural way of implementing the restriction $\xi^2>0$ is via 
\bea  X^{0'} \ = \ \frac{\sqrt{\xi^2}}{\sqrt{M}}\,\cosh(\sqrt{M}\phi) \ , \qquad X^{1} \ = \ \frac{\sqrt{\xi^2}}{\sqrt{M}}\,\sinh(\sqrt{M}\phi) \ ,\label{hyperphi}\eea
where $\phi\in[0,2\pi)$, as a consequence of the identification, and $\sqrt{\xi^2}$ is extracted as the principal square root. In other words, as $\xi^2=0$ is a singularity, the quotient manifold with $\xi^2\geq 0$ is further restricted to the submanifold in which $\xi>0$. Intrinsic Kruskal--Szekeres-like  coordinates are then completed by introducing stereographic $\tx^m$, $m=0,2,3$, such that 
\be X^m \ = \ \frac{2\tx^m}{1-\tx^2} \ , \qquad \xi \ = \ \sqrt{M}\,\frac{1+\tx^2}{1-\tx^2} \ ,\label{remainingKruskal}\ee
where $\tx^m\in\R$ with $-1<\tx^2<1$. The resulting metric takes the form
\be ds^2_{BGM} \ = \ \frac{4d\tx^2}{(1-\tx^2)^2}+\xi^2 d\phi^2 \ . \label{BGMKruskal} \ee
The induced geometry is thus given by the warped product\footnote{We use a notation in which $ds^2_{M\times_f N}=ds^2_M+f^2 ds^2_N$ where $f:M\to \Real$.} ${\rm CMink}_3\times_\xi S^1_K$, where 
\be ds_{\rm CMink_3}^2:=\left. \left(-d\xi^2/M-(dX^0)^2+(dX^2)^2+(dX^3)^2\right)\right|_{-\xi^2/M-(X^0)^2+(X^2)^2+(X^3)^2)=-1}\ ,\ee
is the metric on one of the two stereographic coordinate charts \eq{stereo-embed2} of $AdS_3$. The black hole symmetry group is given by $Stab_{\mso(2,3)}(K)$, i.e., in the realization above chosen, $U(1)_P\times Sp(2)_B$, where the generators of $Sp(2)_B$ are $B\equiv M_{03},M_{02},M_{23}$. 

Just like in the standard BTZ black hole, there is no curvature singularity at $\xi^2=0$, but the spacetime metric \eq{BGMKruskal} evidently degenerates on that surface. Moreover, as a result of the quotient construction, in the spinless case the induced topology of may turn out to be non-Hausdorff at fixed points of $\Gamma$ \cite{BHTZ}. 

However, as we shall recall below, such pathologies are artifacts of the metric-like formulation that are avoided in the unfolded construction. Besides, as a byproduct of the intrinsic unfolded formulation, it will be natural to extend the BGM spacetime beyond the region $\xi=0$.

\paragraph{Intrinsic unfolded construction.}

Instead of starting from identifications in an ambient space in order to produce a black hole, the unfolded constructions of BHTZ-like geometries with topology ${\cal M}_4={\cal M}_3 \times S^1_K$ is obtained by building a factorized gauge function of the type
\be L=\exp_\star(i K\phi)\star \check{L}\ ,\qquad \check{L}:{\cal M}_3\to R(Mp(4;\R))\ ,\label{factorL}\ee
where $K$ is the rigid generator in $\mso(2,3)$ that corresponds to the identification Killing vector $\overrightarrow{K}$, $\phi\in [0,2\pi)$ coordinatizes $S^1_K$, and $\check{L}$ is a gauge function built out of the remaining transvection generators, %strictly periodic on $S^1_K$ 
and subject to conditions at boundaries or other defects of ${\cal M}_4$. 
As a consequence of the factorization of $L$ the flat one-form connection splits as 
\be \O \ = \ L^{\star(-1)}\star dL \ = \ \check{L}^{\star(-1)}\star d\check{L}+i \, \check{L}^{\star(-1)}\star K\star \check{L}\,d\phi \ ,  \ee
reflecting the warped product geometry ${\cal M}_3 \times_\xi S^1_K$, and the resulting holonomy
\be {\rm Hol}_{S^1_K}(\Omega)=\exp_\star(2\pi i K)\ ,\ee
is given by $\gamma$, the generator of $\Gamma$.  

Indeed, the four-dimensional BGM black hole in Kruskal--Szekeres coordinates \eq{BGMKruskal}
can be obtained starting from the gauge function 
\be L_{\rm BGM} \ = \  \exp_\star (iK\phi)\star \exp_\star (i\tm^m P_m )\ , \qquad m=0,2,3 \ ,\label{LBGM}\ee
where 
\be \tm^m(\tx)=4\,{\rm arctanh}\left(\sqrt{\frac{1-\tilde{h}}{1+\tilde{h}}}\right)\,\frac{\tx^m}{\sqrt{\tx^2}}\, \qquad \tilde{h}:=\sqrt{1-\tx^2} \ ,\qquad -1<\tx^2\equiv \tx^m\tx_m<1 \ . \ee 

Recalling, from the discussion leading up to \eq{veryextLstereo}, that for global $AdS_4$ it is possible to move from the covering using two charts, each chart corresponding to a conformal Minkowski spacetime, to a single covering by letting go of the assumption $-1<x^2<1$, it is natural to drop this assumption in \eq{LBGM}, too. This leads to an extension of the BGM black hole obtained by turning the ${\rm CMink_3}$ factor in the BGM geometry into an entire $AdS_3$, thus corresponding to a geometry of type
\be {\rm ExtBGM} \ = \ AdS_3\times_\xi S^1_K \ , \ee
with metric 
\be ds^2_{\rm ExtBGM} \ = \ \frac{4d\tx^2}{(1-\tx^2)^2}+\xi^2 d\phi^2 \ , \qquad \tx^2\neq 1 \ , \label{extBGMKruskal} \ee
In Kruskal-Szekeres coordinates. Clearly, the same extension applies to the gauge function, with an $L_{\rm ExtBGM}$ identical to \eq{LBGM} with $\tx^m \in \R^{1,2}$. 

Note that, via \eq{remainingKruskal}, removing the constraint $-1<\tx^2<1$ amounts to letting $\xi\equiv\sqrt{\xi^2}$ in \eq{hyperphi} take also negative values, i.e., to removing the constraint $\xi>0$ that the BGM spacetime was originally endowed with. In this sense, one could describe the extended BGM manifold via the same embedding \eq{hyperphi}-\eq{remainingKruskal} but taking both signs for the square root of $\xi^2$, i.e.,
\bea & \displaystyle X^{0'} \ = \ \frac{\xi}{\sqrt{M}}\,\cosh(\sqrt{M}\phi) \ , \qquad X^{1} \ = \ \frac{\xi}{\sqrt{M}}\,\sinh(\sqrt{M}\phi) \ ,& \nn \\ 
& \displaystyle X^m \ = \ \frac{2\tx^m}{1-\tx^2} \ , \qquad \xi \ = \ \sqrt{M}\,\frac{1+\tx^2}{1-\tx^2} \ ,  \ \qquad \  \xi\gtreqless 0 \ .\label{extKruskal}\eea
In other words, the above extension of the spinless BGM black hole is obtained by gluing together two ${\rm CMink}_3$ into a (proper) $AdS_3$ across the two surfaces where $\xi$ vanishes.

The extended BGM black hole above was obtained  --- according to Eq. \eq{factorL} --- by separating out a factor, related to the generator of the discrete subgroup determining the identification, from the $AdS$ gauge function $L_{\rm stereo}$; and then applying to the remaining factor the same analytic extension in the coordinates that enabled us to cover the entire AdS manifold as in \eq{veryextLstereo}. Thus, it is natural to expect that, starting by separating out the $K$-dependent factor from the $AdS$ gauge function $L_{\rm spherical}$ \eq{Lspher}, which covers $AdS_4$ entirely, we can obtain the extended BGM spacetime in coordinates in which the warping factor $\xi\gtreqless 0 $ manifestly.

Let us then consider
\be L_{\rm ExtBGM} =  \exp_\star (iK\phi)\star \exp_\star (iET)\star \exp_\star (i \,n^r P_r \,{\rm arcsinh} \rho )\ ,\label{LBTZspher}\ee
where $r=2,3$, $\phi\in [0,2\pi)$, $T\in[0,2\pi)$, $\rho\in\mathbb{R}^+$ and $n^r n_r=1$ parameterize $S^1$. The gauge function is $2\pi$-periodic in $T$, as $\exp_\star (2\pi i E)$ is a central element in $Mp(4;\R)$. As expected, the corresponding $\mathfrak{so}(2,3)$-valued one-form $\Omega=L^{-1}\star dL$ consists of a quasi-frame field $e^a$ and Lorentz connection $\omega^{ab}$ that are bounded and constantly curved, though $e^a$ degenerates at $\xi=0$.
Indeed, this gauge function yields the line element for $AdS_3\times_\xi S^1_K$,
\be ds^2=ds^2_{AdS_3}+\xi^2 d\phi^2 \ , \ee
with $ds^2_{AdS_3}=-(1+\rho^2)dT^2+\frac{d\rho^2}{1+\rho^2}+\rho^2d\psi^2$ and
\be \xi= \sqrt{M}\cos T \sqrt{1+\rho^2} \ \gtreqless \ 0 \ . \ee
This extended spacetime is still a local parametrization of the hyperboloid \eq{hyperb} given by
\bea & X^{0'} \ = \ \sqrt{1+\rho^2}\,\cos T\,\cosh(\sqrt{M}\phi) \ , \qquad X^{1} \ = \ \sqrt{1+\rho^2}\,\cos T\,\sinh(\sqrt{M}\phi) \ , & \nn\\[5pt]
&\displaystyle  X^0 \ = \ \sqrt{1+\rho^2}\,\sin T \ , \qquad X^2 \ = \ \rho \cos\psi \ ,\qquad X^3 \ = \ \rho\sin\psi\ ,&\eea
and can be described as two BGM black holes with $\xi>0$ and $\xi<0$, respectively, glued together across their singularities at $\xi=0$ into a single \emph{topologically extended spinless BGM black hole}\footnote{The closed time-like curves can be removed by going to the covering space of $AdS_3$ leading to four-dimensional geometry with topology $\Real^3\times S^1$.}, with a single conformal infinity. The singularities occur at $T=\pi/2$ and $T=3\pi/2$, where the trapped warped circle shrinks to zero size, and they have $\Real^2\times S^1$ topology and are hidden behind future and past horizons at $\xi=\pm \sqrt{M}$.
Restricting $T$ to $(\pi/2,3\pi/2)$ yields the standard spinless BGM black hole.

Thus, by implementing a specific topology intrinsically, by means of a gauge function, it is natural to extend the BGM manifold beyond the singularity. Besides, note that, not relying explicitly on identifying points along a Killing vector field orbit in an ambient spacetime, the unfolded construction avoids the problem of the quotient BTZ-like manifold being non-Hausdorff where $\vec K$ vanishes. 

However, as $ds^2_{\rm ExtBGM}$ degenerates at $\xi=0$, it remains to be seen how fluctuation fields experience the singularity.

\scs{Fluctuations and resolution of singularities}\label{sec:fluct}

Having explored locally $AdS$ vacua, we shall now turn our attention to the construction of fluctuation fields over them, focussing on the Weyl zero-form sector and comparing metric-like and unfolded approaches. 

As is well-known, in the ordinary, metric-like approach, solution spaces to eqs. \eq{2.30}-\eq{2.32} are built by finding a general solution to the differential equation, which is often simplified by imposing symmetries, and then subjecting it to regularity and boundary conditions.

As explained in Section \ref{sec:unfolding}, studying fluctuations in unfolded approach means solving the linearized twisted adjoint equation in \eq{COMST}, which can be done locally as in \eq{LrotPhi}. Thus, at fixed gauge function, spacetime features distinguishing linearized solutions --- regularity, boundary conditions, etc. --- turn out to be mirrored by algebraic properties of their fibre representative $\Phi'(Y)$.

Sending the reader to the literature for a detailed and extended treatment applied to various noteworthy solution spaces (see \cite{2005,2007,fibre,2011,2012,2015,2017,cosmo,ourBTZ,COMST}, and \cite{Corfu} for a review of the methods), let us briefly recall a few relevant features of the construction starting from how $AdS$ massless particle modes can be encoded into fibre elements $\Phi'(Y)$. 

\paragraph{Example: fibre representatives of massless particle modes.} $AdS$ critically-massless particle modes are solutions to the free field equations with appropriate, spin-dependent mass term, distinguished by regularity conditions in the interior and boundary conditions such that the Killing energy is conserved. The latter condition translates into a quantization of energy, leading to solutions characterized by discrete quantum numbers under the compact subalgebra $\mso(3)\oplus\mso(2)$ generators \cite{Breitenlohner:1982jf,Mezincescu:1984ev}. Stripping off the spacetime dependence by virtue of \eq{LrotPhi} and \eq{Lkill}, the latter condition can be imposed algebraically on $\Phi'(Y)$, via the twisted-adjoint action of the $AdS$ isometry generators $E$ and $M_{rs}$: %
\bea
&[E,\Phi']_\pi=\{E,\Phi'\}_\star  =e\,\Phi'\,,&\\ 
&\frac12 [M^{rs},[M_{rs},\Phi']_\pi]_\pi =\frac12 [M^{rs},[M_{rs},\Phi']_\star]_\star =s(s+1)\,\Phi'\,,&
\eea
where the second condition fixes the eigenvalue of the quadratic Casimir $\frac12 M^{rs}\star M_{rs}$ of $\mso(3)$. Solving these conditions determines $\Phi'=T_{e;(s)}$, a $(2s+1)$-plet of non-polyomial functions in $Y$, with elements distinguished by the eigenvalue $j_s$ of one specific spatial rotation $J$ (say $J=M_{12}$) $j_s = -s, -s+1,\ldots, s-1, s $. $Y$-space elements like $T_{e;(s);j_s}$ are thus fibre counterparts of particle modes, and span lowest-weight  modules (highest-weight modules for the anti-particle states) built via the action of energy-raising (lowering) operators $L^+_r$ ($L^-_r$) on a lowest-weight (highest-weight) state $T_{e_0;(s_0)}$ ($T_{-e_0;(s_0)}$), singled out by \cite{fibre}
\be [L^-_r,  T_{e_0;(s_0)}]_\pi =L^-_r \star  T_{e_0;(s_0)}-T_{e_0;(s_0)}\star L^+_r =0
\,,\qquad{\rm for}\quad 
e_0 = s_0+1 
\,.
\label{eq:LWcond}\ee
For example, the ground state of the $AdS$ massless scalar particle with pure Neumann boundary conditions has energy eigenvalue $e_0=1$, i.e., is singled out by the conditions 
\be [M_{rs},\Phi']_\star \ = \ 0 \ , \qquad   \{E,\Phi'\}_\star \ = \ \Phi' \ ,\ee
which are solved by
\be \Phi' \ =  \ T_{1; (0)} \ = \ 4e^{-4E} \ ,\label{ELWS} \ee
which indeed solves the lowest-weight condition \eq{eq:LWcond}. The fact that this fibre element indeed corresponds to the regular solution of the $AdS$-massless Klein-Gordon equation \eq{2.30}, can be checked by reinstating the spacetime dependence via \eq{LrotPhi} using a gauge function. For instance, in spherical coordinates,
\bea \Phi_{1;(0)} \ = \ L_{\rm spherical}^{-1}\star T_{1;(0)}\star \pi(L_{\rm spherical}) \ = \ \frac{e^{-it}}{\sqrt{1+r^2}}\,e^{iyM\yb} \ ,\label{PhiBF} \eea
where the spacetime-dependent matrix $M_{\a}{}^{\bd}$ is given in \cite{2017,Corfu}.
Thus, the $y=0=\yb$ component of the master field \eq{PhiBF} is, as expected, the ground state scalar field with pure Neumann boundary conditions,
\be C(t,r) \ = \ \frac{e^{-it}}{\sqrt{1+r^2}}  \ , \ee
and all the higher modes are stored in the $Y$-expansion of $\Phi_{1;(0)}$.

\paragraph{Cartan bases and fibre operator algebras.}   As explained in detail in \cite{fibre}, the elements $T_{e;(s)}$, with definite eigenvalues under the compact subalgebra $\mso(2)_E\oplus \mso(3)_{M_{rs}}$ of $\mso(2,3)$, provide basis elements for a fibre dual of the standard harmonic expansion of the Fronsdal fields. 
More generally, one may expand the zero-form integration constants using fibre operators with definite eigenvalue under different, non-compact subalgebras $\mso(1,1)\oplus\mso(1,2)$ of $\mso(2,3)$, which can be considered as non-compact fibre duals of generalized harmonics corresponding to alternative boundary conditions in spacetime. As described in \cite{Corfu}, these generalized fibre harmonics can be realized starting from a class of operators realizing Fock-space endomorphisms and built starting from a choice of two elements $K_{(\pm)}$ in the Cartan subalgebra of the complexified $AdS$ isometry algebra $\msp(4;\C)$, with oscillator realization 
\be K_{(\pm)}=\frac{1}{8} K^{(\pm)}_{\underline{\a\b}}Y^{\underline\a}\star Y^{\underline\b}\ ,\ee
where ($q=\pm$)
\bea [K^{(q)},K^{(q')}]_{\au\bu}~=~0\ ,\qquad  K^{(q)}_{\au}{}^{\underline{\gamma}}\,K^{(q)}_{\underline{\gamma}}{}^{\bu}~=~-\delta_{\au}{}^{\bu}\ .\label{K2}
\eea
By virtue of these properties, the chosen elements $K_{(\pm)}$ can be used to split the $Y$ oscillators into two sets of creation/annihilation operators $(a^+_i,a^-_i)$ ($i=1,2$) with Weyl-ordered number operators $w_i=a^+_ia^-_i$ (no sum over $i$) such that
\be K_{(\pm)}=\frac{1}{2}(w_2\pm w_1)\ .\ee
It is then possible to build operators $P_{{\mathbf n}_L|{\mathbf n}_R}(Y)$ obeying 
\be P_{{\mathbf n}_L|{\mathbf n}_R}=\pi\bar\pi(P_{{\mathbf n}_L|{\mathbf n}_R})\ ,\ee
and 
\be P_{{\mathbf n}_L|{\mathbf n}_R} \star P_{{\mathbf m}_L|{\mathbf m}_R} \ = \ \delta_{{\mathbf n}_R,{\mathbf m}_L} P_{{\mathbf n}_L|{\mathbf m}_R} \ , \label{projalg}\ee
with ${\bf n}_{L,R}=(n_1,n_2)_{L,R}\in ({\mathbb Z}+1/2)\times ({\mathbb Z}+1/2)$, {\it idem} ${\bf m}_{L,R}$, being half-integer eigenvalues under the left or right star-product action of number operators $w_i$,
\be (w_i-n_{iL}) \star  P_{{\mathbf n}_L|{\mathbf n}_R} \ = \ 0 \ = \ P_{{\mathbf n}_L|{\mathbf n}_R} \star (w_i-n_{iR}) \ .\label{LReigen} \ee
Clearly, the $P_{{\mathbf n}_L,{\mathbf n}_R} $ also diagonalize the adjoint as well as twisted-adjoint actions of $K_{(\pm)}$, \emph{viz.} 
\be K_{(\pm)}\star P_{{\mathbf n}_L|{\mathbf n}_R} -P_{{\mathbf n}_L|{\mathbf n}_R}  \star K_{(\pm)} \ = \ \frac12\left(n_{2L}\pm n_{1L} -(n_{2R}\pm n_{1R}) \right)P_{{\mathbf n}_L|{\mathbf n}_R} \ ,\label{adjeig} \ee
\be K_{(\pm)}\star P_{{\mathbf n}_L|{\mathbf n}_R} -P_{{\mathbf n}_L|{\mathbf n}_R}  \star \pi(K_{(\pm)}) \ = \ \frac12 \left(n_{2L}\pm n_{1L} -(-1)^{\sigma_\pi(K_{(\pm)})} (n_{2R}\pm n_{1R}) \right)P_{{\mathbf n}_L|{\mathbf n}_R} \ ,\label{twadjeig} \ee
where $\pi(K_{(\pm)}) = \sigma_\pi(K_{(\pm)}) K_{(\pm)}$. 

Adjoint and twisted-adjoint action \eq{adjeig}-\eq{twadjeig} are only different for $\pi$-odd Cartan generators (i.e., transvections). Since $K_{(\pm)}\star\kappa_y=\kappa_y\star\pi(K_{(\pm)})$, for any $\pi$-odd $K_{(\pm)}$ star-multiplication of $P_{{\mathbf n}_L|{\mathbf n}_R}$ by $\kappa_y$ exchanges adjoint and twisted-adjoint action, e.g. $K_{(\pm)}\star P_{{\mathbf n}_L|{\mathbf n}_R}\star\kappa_y-P_{{\mathbf n}_L|{\mathbf n}_R}\star\kappa_y\star\pi(K_{(\pm)}) = [K_{(\pm)},P_{{\mathbf n}_L|{\mathbf n}_R}]_\star \star\kappa_y$ .  This means, in particular, that in such cases one can define operators that mix Fock and anti-Fock space states, or \emph{twisted operators}, via star-multiplication by $\kappa_y$: as $\kappa_y\star w_1 = -w_2\star \kappa_y$,  a twisted operator $P_{{\mathbf n}_L|{\mathbf n}_R}\star \kappa_y$ has right eigenvalues
\be \tP_{{\mathbf n}_L|{\mathbf n}_R} \ := \ P_{{\mathbf n}_L|{\mathbf n}_R}\star \kappa_y \ \sim \  P_{{\mathbf n}_L|-n_{2R},-n_{1R}} \ .\ee
Like $\kappa_y$ itself, such twisted counterparts of the Fock space endomorphisms are distributions in $Y$ in Weyl order, whereas the $P_{{\mathbf n}_L|{\mathbf n}_R}$ are regular \cite{2017,COMST,Corfu}. 

There are three distinct choices of $(K_{(+)},K_{(-)})$ modulo $Sp(4;\Real)$ rotations, corresponding to pairs of commuting compact, non-compact or mixed generators. With a conventional choice of such generators, the three pairs can be given by \cite{2011}
\be (E,J) \ , \qquad (J,iB) \ , \qquad (iB,iP) \ ,\ee
where $E:=P_0=M_{0'0}$ is the AdS energy, $J:=M_{12}$ is a spin, $B:= M_{03}$ is a boost and $P:=P_1=M_{0'1}$ is a transvection. Thus, starting from a pair of Cartan generators, one may form four lowest-weight ($\e=-$) or highest-weight ($\e=+$) projectors, namely $\exp(4\e K_{(\e')})$, where $\e,\e'=\pm$, and their twisted counterparts $\exp(4\e K_{(\e')})\star \kappa_y$, which are distinct elements iff $K_{(\e')}=E$ or $iP$ (as $\exp(\pm 4J)\star \kappa_y=\exp(\pm 4J)$, \emph{idem} $iB$).
Once a pair is chosen, then the orbit of a chosen $\exp(4\e K_{(\e')})$ and twisted counterpart under the left and right actions of ${\cal H}$  form an associative algebra ${\cal M}_\e(K_{(\e')};K_{(-\e')})$, with \emph{principal Cartan generator} $K_{(\e')}$.
Letting ${\cal M}(K_{(\e')};K_{(-\e')})={\cal M}_+(K_{(\e')};K_{(-\e')})\oplus {\cal M}_-(K_{(\e')};K_{(-\e')})$, we thus have \emph{six} possibilities,
\be {\cal M}(E;J)\ ,\quad {\cal M}(J;E)\ ;\qquad {\cal M}(J;iB)\ ,\quad {\cal M}(iB;J)\ ;\qquad {\cal M}(iB;iP)\ ,\quad {\cal M}(iP;iB)\label{families} \ .\ee

The (anti-)particle states are obtained, via \eq{LrotPhi} from fibre representatives  $\Phi'\in{\cal M}(E;J)$, while master fields built from elements $\Phi' \in {\cal M}(iB;iP)$ and $\Phi' \in{\cal M}(iP;iB)$ are of relevance for the unfolded analysis of fluctuations over the ${\rm BGM}$ and ${\rm ExtBGM}$ spacetimes, and we shall focus on them in Section \ref{Sec:deg}. 

\paragraph{Regular presentation.} The solution to the eigenvalue equation \eq{LReigen} can be written as 
\be  P_{{\mathbf n}_L|{\mathbf n}_R}=f_{n_{1L}|n_{1R}}(a^+_1,a^-_1)f_{n_{2L}|n_{2R}}(a^+_2,a^-_2)\ ,\label{Pff}\ee
with each factor of the form 
\be f_{n_L|n_R}\left( a^{+},a^{-}\right) \ = \ \mathrm{C}_{n_L,n_R}\,\left( a^{+}\right)
^{n_{L}-n_{R}}e^{-2w}L_{ n_{R}-\frac{1}{2}}^{n_{L}-n{R}}\left( 4w\right)  \ ,\label{fexp} \ee
where $\mathrm{C}_{n_L,n_R}$ is a normalization constant and $L_{ n_{R}-\frac{1}{2}}^{n_{L}-n_{R}}$ a generalized Laguerre polynomial \cite{2011}. Strictly speaking, the above form holds for positive half-integers $n_{L,R}$ such that $n_L\geq n_R$. However, it admits analytic continuation to the case when $n_{L}, n_R$ become in fact complex numbers $\l_L,\l_R$ \cite{ourBTZ}. The reason for considering this extension is that, in building fluctuations over the ${\rm (Ext)BGM}$ black hole, the \emph{imaginary} part of the eigenvalues of $\Phi'$ under $K_{(-)}=\frac{w_2-w_1}{2}=iP$ will be quantized in order to be periodic on the $S^1_K$ cycle. 

In order to perform star product calculations we shall therefore endow our basis fibre elements $f_{\l_L|\l_R}$ with a specific integral presentation in terms of Gaussian functions \cite{2011,2017,ourBTZ,COMST}, prescribing to perform all star products and traces prior to computing the auxiliary integrals. We shall refer to this scheme as \emph{regular presentation} of the master fields. This scheme is also crucial for the simple case of diagonal operators (projectors) $P_{{\mathbf n}|{\mathbf n}}$ with generic half-integer ${\mathbf n}=(n_1,n_2)$, in order to remove potential divergencies in the star product between states with positive and negative $K_{(q)}$ eigenvalue \cite{2017,Corfu}, thereby obtaining a concrete realization of the associative algebra \eq{projalg} fully extended to ${\bf n}_{L,R}\in ({\mathbb Z}+1/2)\times ({\mathbb Z}+1/2)$. 

The simplest regular presentation that satisfies the above requirements is 
\be f_{\lambda _{L}|\lambda _{R}}\left( a^{+},a^{-}\right) \ = \ {\cal N}_{\lambda _{L},\lambda _{R}}%
\int_{0}^{+\infty }d\tau \frac{\tau ^{\lambda _{R}-\lambda _{L}-1}}{\Gamma
\left( \lambda _{R}-\lambda _{L}\right) }e^{-\tau a^{+}}\ \oint_{C(\pm1)}\frac{%
d\varsigma }{2\pi i}\ \frac{\left( \varsigma+1 
\right) ^{\lambda _{L}-\frac{1}{2}}}{\left( \varsigma -1\right) ^{\lambda
_{R}+\frac{1}{2}}}\, e^{-2\varsigma w}\text{ ,}  \label{solnint} \ee
where ${\cal N}_{\lambda _{L},\lambda _{R}}$ is a normalization, the first integral is a Mellin transform which helps extending the factor $(a^+)^{\l_L-\l_R}$ to complex $\l_{L,R}$ \footnote{Strictly speaking, the integral $\left( a^{+}\right) ^{\lambda _{L}-\lambda _{R}}=\int_{0}^{+\infty }d\tau 
\frac{\tau ^{\lambda _{R}-\lambda _{L}-1}}{\Gamma \left( \lambda
_{R}-\lambda _{L}\right) }e^{-\tau a^{+}}$ only makes sense for $\mathrm{Re}\left( \lambda _{L}-\lambda_{R}\right) <0$ 
and $\mathrm{Re}\left( a^{+}\right) > 0$. In order to extend it to any $\lambda _{L}-\lambda _{R} \neq - 1,- 2,...$ and $\mathrm{Re}( a^{+}) >0$, we can analytically continue it with 
$\left( a^{+}\right)^{\lambda _{L}-\lambda _{R}} \ = \ \Gamma(1+\lambda _{L}-\lambda _{R})\int_\gamma \frac{d\tau}{2\pi i}\,{\tau^{\lambda _{R}-\lambda _{L}-1}}\,e^{\tau a^+} 
$, where $\gamma$ is a contour of Hankel type \cite{ourBTZ}. In practice, when evaluating the spacetime-dependent master field it will be possible to formally use the simpler presentation included in \eq{solnint}, and then analytically continue $\lambda _{L}-\lambda _{R}$ beyond the region $\mathrm{Re}\left( \lambda _{L}-\lambda_{R}\right) <0 $ after all star-products have been evaluated.}, $\Gamma $ is the gamma function, and the second integral is a closed contour Laplace transform encoding the remaining, $w$-dependent factors in \eq{fexp}. More precisely, $C(\pm 1)$ is a small closed contour encircling $\pm 1$, and in order for it not to cross any branch cut of the integrand we shall work with the limitation 
that 
\bea &\l_L\in\C \ , \quad \lambda _{R}+\frac{1}{2}\ \in \ \mathbb{ Z}^{+}  \ , \quad {\rm for} \ \ \varsigma_0 = 1& \ ,\label{lpmL}\\ 
&\lambda_{L}-\frac{1}{2}\ \in \ \mathbb{ Z}^{-}  \ ,  \quad \l_R\in\C \ , \quad {\rm for} \ \ \varsigma_0 = -1 &\ .\label{lpmR}\eea
which are sufficient for a first analysis of fluctuations over the (Ext)BGM background\footnote{More general integral presentations that forego this limitation are explored in \cite{Yihao}.}. For the sake of simplicity of the regular presentation, we shall consider expanding our fluctuation fields only over eigenfunctions of type \eq{lpmL} --- which implies that condition \eq{lpmR} features in their hermitian conjugates, that the reality conditions require \cite{ourBTZ}. Finally, one can check that in the limit $\l_{L}-\l_{R}\to 0$, the integral presentation \eq{solnint} of $f_{\l_L|\l_R}$ smoothly reduces to that of an ordinary Fock-space projector $f_{\l_R|\l_R}$ 
\begin{eqnarray}
&&f_{\lambda _{L}|\lambda _{R}}\left( a^{+},a^{-}\right) \quad \xrightarrow[\l_{L}-\l_{R}\to 0]{} \quad  \oint_{C(\varepsilon)}\frac{%
d\varsigma }{2\pi i}\ \frac{\left( \varsigma+1 
\right) ^{\lambda _{R}-\frac{1}{2}}}{\left( \varsigma -1\right) ^{\lambda
_{R}+\frac{1}{2}}}\, e^{-2\varsigma w}\text{ ,}\label{diaglim}
\end{eqnarray}
where now $\varepsilon={\rm sign}(\l_R)$. The product of two such projectors, according to \eq{Pff}, with $n_{iL}=n_{iR}=\frac12$, within ${\cal M}(E,J)$ gives rise to the regular presentation of the scalar particle ground state \eq{ELWS} \cite{2011,2017,COMST,Corfu}. 

While elements like $f_{\l_L|\l_R}$ are in general non-analytic in $Y$ for $\l_L \in \Comp$, and thus incompatible with a physical interpretation of the expansion coefficients in terms of fields of various spins, reinstating the spacetime dependence via the gauge function $L$ in fact removes this problem except at singularities (provided  that  the  star  products  with $L$ are performed prior to taking the limit back to the unfolding point). 

\vspace{0.5cm}

Below, we shall apply the above formalism to construct fluctuations over the BGM background. We shall see how, due to the spacetime/fibre duality, possible singularities of individual fields will acquire a more transparent meaning at the level of the master field, which, to a large extent (to be reviewed below), remains in fact smooth.
To this end, it will be useful to first briefly review how Schwarzschild-like curvature singularities, appearing in the four-dimensional spherically symmetric higher-spin black hole solutions, are resolved in the sense above described.

\scss{Resolution of curvature singularities}

This subject has been treated in detail in \cite{2011,2017,ourBTZ,Corfu}, so we shall here only recall the basic idea, that will be of relevance for the following.  The full Vasiliev equations admit higher-spin black-hole-like solutions \cite{Didenko:2009td,2011,2012,2017,COMST}, obtained from twisted projectors $\tP_{\mathbf{n}|\mathbf{n}}$ in the family ${\cal M}(E,J)$. In the spherically-symmetric case the Weyl zero-form contains, as coefficients of their $Y$-expansion, a tower of type-D spin-$s$ Weyl tensors of the form
\be \Phi_{{\textrm{bh}},\a(2s)}\ \sim \ \frac{\nu}{r^{s+1}}\,(u_{(E)}^+
 u_{(E)}^- )^s_{\a(2s)} \label{nsWeylspher}  \ee
(together with their analogues for the anti-selfdual part), where $u_{(E)}^\pm$ are the principal spinors. The spin-$2$ Weyl tensor coincides with that of an $AdS_4$ Schwarzschild black hole. Each individual generalized Weyl tensor (including the spin $s=0$ and $s=1$ elements) correspond to static, singular solutions of the corresponding spin-$s$ free Klein-Gordon, Maxwell, and Bargmann-Wigner equations \cite{Didenko:2009td,Corfu}, and is evidently divergent in $r=0$. 

Note that, in the simplest examples of such solutions, the deformation parameter that turns on the entire solution, $\n$ in \eq{nsWeylspher}, is independent of $s$, i.e. it is the same for the entire tower of Weyl tensors. As that parameter is connected to spin-$s$ asymptotic charges \cite{Didenko:2015pjo} (see however some caveats with this interpretation \cite{universe,COMST}), this manifests a sort of extremality of such solutions --- which one can forego by building a higher-spin black hole via a sum over an ensemble of solutions with the same asymptotics  \cite{2011,2012,2017,COMST}. 

While a proper analysis of the singularity requires a higher-spin extension of the ordinary concepts of Riemannian geometry, such as a higher-spin invariant generalization of the line element, it is interesting to observe how the higher-spin embedding of the ordinary gravitational black hole immediately renders the singularity more tractable. Indeed, the divergencies of the individual spin-$s$ curvatures acquire a clearer meaning for the higher-spin covariant master field $\Phi_{\rm bh}(x,Y)$, which gives rise to the Weyl tensor generating function
\be \displaystyle \left.\Phi_{\rm bh} \right|_{\bar y=0} \ \propto \
\frac{1}{r}\,\exp \left(\ft{1}{2r}\,
y^\a{\cal D}^{(E)}_{\a\b}y^\b\right)\ ,\label{o3Weylnm} \ee
out of which \eq{nsWeylspher} are extracted (${\cal D}^{(E)}_{\a\b}=u_{(E)\a}^+
 u_{(E)\b}^-+u_{(E)\a}^-
 u_{(E)\b}^+$). Eq. \eq{o3Weylnm} is in fact a delta-sequence in $y$ with $r$ playing the role of the $\e$-parameter, i.e.,
 \be \left.\Phi_{\rm bh} \right|_{\bar y=0} \quad \xrightarrow[r\to 0] \ \ \ \ \   2\pi\d^2(y) \ ;  \ee
in other words, the individual singularities of the Weyl tensors assemble into a distributional fibre behaviour for their generating function. However, this mapping makes the problem more transparent and tractable, since a delta function of non-commutative variables can be considered smooth as it is well-behaved under star product \cite{2017,cosmo,COMST,meta}. As stressed in \cite{meta}, delta functions of non-commutative variables are in fact equivalent to bounded functions up to a change in the ordering prescription: as these leave invariant the classical observables \cite{2005,Sezgin:2011hq,universe,COMST} of the Vasiliev system (possibly up to subtle boundary terms in oscillator space) the resolution of such curvature singularities would amount to declaring them artifacts of the ordering choice for the infinite-dimensional symmetry algebra governing the Vasiliev system.
 
So at curvature singularities of this kind the component field picture breaks down, but the differential graded algebra of master fields is still well defined. It is in this sense that we say that the higher-spin embedding resolves the singularity in $r=0$ of the spherically-symmetric black hole. 

We shall see in the following that a similar singularity also appears in fluctuations over the (Ext)BGM spacetime, and can be resolved by a similar mechanism. 

%r=0: AdS particles and bhs in HS embedding, curvature singularities turned into smooth fibre space delta function elements.

%Stress the regular prescription + NC delta function to resolve the singularity --- all spins are involved!

%-- whether xi=0 is singular or not s¿we ask by using fluctuations as out probe...

%Switch on the fluctuation of cite{2019}. Explain table, possibly draw parallels with ordinary solution of KG on BTZ. 

%xi=0 is resolved here one only gets the phenomena of rapid oscillations in ''time'' (divergent locally defined frequency)

\scss{Degenerate metrics}\label{Sec:deg}

\paragraph{Scalar field on BGM black hole in metric-like approach.}

In order to study the behaviour of a (critically) massless scalar field with definite eigenvalues under the action of the two commuting Killing vector fields $\vec{v}^L_{0'1}$ and $\vec{v}^L_{03}$, over a BGM background, and in particular close to the singularity in $\xi=0$, it is convenient to refer to an adapted coordinate system, 
\bea & X^{0'} \ = \ \frac{\xi}{\sqrt{M}}\,\cosh(\sqrt{M}\phi) \ , \qquad X^{1} \ = \ \frac{\xi}{\sqrt{M}}\,\sinh(\sqrt{M}\phi) \ , \qquad X^2 \ = \ x \ ,  & \nn\\[5pt]
&\displaystyle \qquad  X^0 \ = \ \sqrt{1+x^2-\frac{\xi^2}{M}}\,\cosh\gamma \ , \qquad X^3 \ = \ \sqrt{1+x^2-\frac{\xi^2}{M}}\,\sinh\gamma \ ,&\label{xixcoords}\eea
with $\xi,x,\gamma\in \mathbb{R}$ and such that $x^2-\frac{\xi^2}{M}=:-\D^2>0$, $\phi\in[0,2\pi)$. With this parameterization,  $\vec{v}^L_{0'1}=\frac{1}{\sqrt{M}}\frac{\partial}{\partial \phi}$ and $\vec{v}^L_{03}=\frac{\partial}{\partial \gamma}$. 

Introducing variables $\a,\b$ such that $\a^2=M\,\frac{x^2}{\xi^2}$, $\b^2=x^2-\frac{\xi^2}{M}=-\D^2$, $1+\b^2=(\vec{v}^L_{03})^2$, where $\a \in \mathbb{R}$, $\a>1$, and $\b\in\mathbb{R}$, one can rewrite
\bea & X^{0'} \ = \ \frac{\b}{\sqrt{\a^2-1}}\,\cosh(\sqrt{M}\phi) \ , \qquad X^{1} \ = \ \frac{\b}{\sqrt{\a^2-1}}\,\sinh(\sqrt{M}\phi) \ , \qquad X^2 \ = \ \frac{\a\b}{\sqrt{\a^2-1}} \ ,  & \nn\\[5pt]
&\displaystyle \qquad  X^0 \ = \ \sqrt{1+\b^2}\,\cosh\gamma \ , \qquad X^3 \ = \ \sqrt{1+\b^2}\,\sinh\gamma \ ,&\eea
with the further advantage that the metric in these variables is diagonal,
\be ds^2_{\rm ExtBGM} \ = \ \frac{d\b^2}{1+\b^2}-\frac{\b^2}{(\a^2-1)^2}\,d\a^2+(1+\b^2)d\gamma^2+M\frac{\b^2}{\a^2-1}\,d\phi^2 \ . \ee
Considering the scalar field on this background, one has 
\bea \nabla^2 C & = & \frac{1}{\b^2}\,\partial_\b\left(\b^2(1+\b^2)\partial_\b C\right)-\frac{(\a^2-1)^{3/2}}{\b^2}\,\partial_\a\left((\a^2-1)^{1/2}\partial_\a C\right) \nn\\
&& +\frac{\a^2-1}{M\b^2}\partial^2_\phi C+\frac{1}{1+\b^2}\,\partial^2_\gamma C \ . \label{boxab}\eea
Imposing periodicity in $\phi$ and a definite eigenvalue under $\vec{v}^L_{iB}$, a natural Ansatz for the scalar field on the extended BGM spacetime is
\be C(\a,\b,\phi,\gamma) \ = \ e^{in\phi}\,e^{im\gamma}\,f_{nm}(\a,\b) \ .\label{scalaransatz} \ee
Inserting it into the Klein-Gordon equation $(\nabla^2 +2)C=0$ and using \eq{boxab}, we obtain an equation for the function $f_{nm}$ of the form 
\bea  & \displaystyle \frac{1}{\b^2}\,\partial_\b\left(\b^2(1+\b^2)\partial_\b f_{nm}\right)-\frac{(\a^2-1)^{3/2}}{\b^2}\,\partial_\a\left((\a^2-1)^{1/2}\partial_\a f_{nm}\right) &\nn\\
&\displaystyle  \hspace{2cm} +\left(2-\frac{\a^2-1}{M\b^2}\,n^2 -\frac{1}{1+\b^2}\,m^2\right)f_{nm} \ = \ 0  &\ .\eea
This equation can be solved by separating variables as 
\be f^\l_{nm}(\a,\b) \ = \ u^\l_n(\a)\,v^\l_m(\b) \ , \ee
where $\l$ is a separation constant, such that
\bea 
\left[-(\a^2-1)^{3/2}\,\partial_\a\left((\a^2-1)^{1/2}\partial_\a\right)-(\a^2-1)\,\frac{n^2}{M}\right]u^\l_n & = & \l u_n^\l \ , \label{alphaeq}\\ 
\left[\partial_\b\left(\b^2(1+\b^2)\partial_\b\right)-\frac{\b^2}{1+\b^2}\,m^2+2\b^2\right] v_m^\l & = & -\l \,v_m^\l \ .\label{betaeq}\eea
Let us study the case $\l=0$. The general solution of \eq{alphaeq} with $\l=0$ is
\be u^{\l=0}_n(\a) \ = \ c_1\,\cos\left[\frac{n}{\sqrt{M}}\,{\rm arctanh}\left(\frac{\a}{\sqrt{\a^2-1}}\right)\right]+c_2 \,\sin\left[\frac{n}{\sqrt{M}}\,{\rm arctanh}\left(\frac{\a}{\sqrt{\a^2-1}}\right)\right]\ ,\ee
while \eq{betaeq} determines $v^{\l=0}_m$ as
\bea v^{\l=0}_m(\b) & = &  (1+\b^2)^{-im/2}\left[c_3\,{}_2F_1\left(\frac{1-im}{2},\frac{2-im}{2};\frac34;-\b^2\right)\right. \nn\\
&& \left.+c_4\,{}_2F_1\left(-\frac{im}{2},\frac{1-im}{2};\frac12;-\b^2\right)\right]\ ,\eea
where $c_i$, $i=1,2,3,4$ are integration constants. In the following we will be interested in the case $m=0$, which simplifies to
\be v^{\l=0}_0(\b) \ = \ \frac{c_3+c_4\,\arctan \b}{\b} \ .\ee
Thus, in terms of the variables \eq{xixcoords}, for $\l=0=m$
\bea & \displaystyle C(\xi,x,\phi,\gamma) \ = \ \frac{e^{in\phi}}{\sqrt{-\D^2}}\left[c_1\,\cos\left(\frac{n}{\sqrt{M}}\,{\rm arctanh}\frac{x}{\sqrt{-\D^2}}\right)\right.& \nn \\
& \displaystyle \hspace{2cm} +c_2 \,\left.\sin\left(\frac{n}{\sqrt{M}}\,{\rm arctanh}\frac{x}{\sqrt{-\D^2}}\right)\right]\left(c_3+c_4\,\arctan(\sqrt{-\D^2})\right) \ .&  \label{solmetric}\eea

The scalar field diverges at the surface $\D^2\equiv\frac{\xi^2}{M}-x^2=0$, while it remains bounded but oscillates with infinite frequency (as $\frac{x}{\sqrt{-\D^2}}\simeq 1+\frac12\frac{\xi^2}{Mx^2}$, for $\xi\to 0$) at $\xi=0$, $x\neq 0$. In this sense, scalar fluctuations do experience the BGM singularity as a pathological surface.

\paragraph{Fluctuations on (Ext)BGM black hole in unfolded approach.} 
Let us now turn to describing how the above results are recovered in terms of master fields and what conclusions can be drawn about the BGM singularity and the extended BGM manifold from the unfolded approach. 

In order to reproduce a solution like \eq{solmetric}, we shall expand $\Phi'$ over basis fibre eigenfunctions belonging to the extension of the families ${\cal M}(iB,iP)$ or ${\cal M}(iP,iB)$ obtained by acting on their ground states with suitable complex powers of creation and annihilation operators. As anticipated, this will be crucial to non-trivially satisfy the periodicity condition around the $S^1_K$ circle. 

Having chosen $K\propto P$ as generator for the identification, and denoting with $\widetilde{K}$ the commuting generator $B$, such requirements select two linearized moduli spaces with distinct characteristics, given by the unbroken symmetry $H$ and singularity structure of the physical scalar field $C$ of the corresponding ground states, via the following steps:

\begin{enumerate}

\item First, we can either choose the fibre representative of the ground state $\Phi'_0\equiv \Psi'_0\star\kappa_y$ to be in ${\cal M}(iB,iP)$ or ${\cal M}(iP,iB)$, corresponding to a choice of which between $iP $ and $iB$ is the principal Cartan generator. With $iB$ as principal Cartan generator, we can in principle choose whether to expand $\Phi'$ within the regular or the twisted sector, as $\exp(\pm 4iB)$ is an eigenstate of $\kappa_y$. However, sticking to the regular presentation \eq{solnint} lifts the ambiguity, as only an expansion over the twisted sector gives rise to well-defined integrals after reinstating the $x$-dependence via the gauge function (see Appendix E in \cite{ourBTZ}).

\item Then, we should examine which choices are compatible with the identification. 

\end{enumerate}

This leaves only two possible choices for $\Psi'_0$: $\Psi'_0=e^{\pm 4iP}$ or $\Psi'_0=e^{\pm 4iB}$, leading to scalar fields with singularities respectively at the BGM horizon and at the surface $\tilde{\xi}^2=1$, i.e. $\D^2=0$, passing through the BGM horizon and singularity. 

Had we instead chosen $K\propto B$ as generator for the identification, and denoting with $\widetilde{K}$ the commuting generator $P$, the same steps leave three possible choices for $\Psi'_0$: the former two as well as $\Psi'_0=e^{\pm 4iP}\star\kappa_y$, which lead to a scalar field blowing up at the $\tilde{\xi}^2=0$ surface, another membrane-like singularity outside the BGM horizon. 
These results are summarized in Table \ref{Table}\footnote{This table corrects an error in \cite{ourBTZ}, which incorrectly includes $\Psi'_0=e^{\pm 4iP}\star\kappa_y$ in the list of possible fluctuation fields when $iP$ is the identification generator.}.

\begin{table}
\centering
\begin{tabular}{| l || c | c | c |}
\hline
$(K;\widetilde K)$ & $\Psi'_0$ & $H$ & $C$ \\
\hline
$(P;B)$ & $e^{\pm 4iP}$ & $U(1)_P \times Sp(2)_B$ & $\frac{1}{\sqrt{1-\xi^2}}$\\
 & $e^{\pm 4iB}$ & $U(1)_P \times U(1)_B$ & $\frac{1}{\sqrt{1-\widetilde\xi^2}}$\\
 \hline
$(B;P)$ & $e^{\pm 4iP}$ & $U(1)_B \times U(1)_P$ &  $\frac{1}{\sqrt{1-\widetilde\xi^2}}$\\
& $e^{\pm 4iP}\star\kappa_y$ & $U(1)_B$ & $\frac{X^{0'}+X^1}{\widetilde \xi^2}$\\
& $e^{\pm 4iB}$& $U(1)_B \times Sp(2)_P$ & $\frac{1}{\sqrt{1-\xi^2}}$\\
    \hline
    \end{tabular}
\caption{\textbf{Ground states for fluctuations spaces on spinless BGM black holes.} $\protect\overrightarrow K$ and $\protect\overrightarrow{\widetilde K}$, respectively, denote the identification Killing vector and its dual of a (Ext)BGM black hole with mass $M=1$ and spin $J=0$.
The black hole symmetry group is given by $Stab_{\mso(2,3)}(K)$, \emph{i.e.} $Stab_{\mso(2,3)}(P)=U(1)_P \times Sp(2)_B$ and $Stab_{\mso(2,3)}(B)=U(1)_B \times Sp(2)_P$, which is also the stabilizer of the warp factor $\xi:= \sqrt{\protect\overrightarrow K^2}$.
$H$ and $C$, respectively, denote the symmetry group and scalar field of the ground state $\Psi_0$ of a sector of fluctuations.
%
%There are four distinct moduli spaces, depending on whether $C$ blows up at 1) $\xi=0$, \emph{i.e.} at the BGM singularity; 2) $\xi=\pm 1$, \emph{i.e.} at the BGM horizons; 3) $\widetilde \xi:= \sqrt{\protect\overrightarrow{\widetilde K}^2}=0$, \emph{i.e.} at a membrane-like singularity outside the BGM horizons; and 4) $\widetilde \xi=1$ (also denoted by $\Delta=0$), \emph{i.e.} at a membrane-like singularity passing through the BGM horizon and singularity.
}
\label{Table}\end{table}

In what follows we shall focus on $iB$ as principal Cartan generator, and we shall expand $\Psi'$ on eigenfunctions of the form 
\be  f_{\boldsymbol{\lambda }}\left( a_{1}^\pm ,a_{2}^\pm \right) \ := \ f_{\boldsymbol{\lambda }_1}\left(a_{1}^\pm\right) f_{\boldsymbol{\lambda }_2}\left( a_{2}^\pm\right) :=f_{\l_{1L}|\l_{1R}}(a^+_1,a^-_1)f_{\l_{2L}|\l_{2R}}(a^+_2,a^-_2)\ ,\label{Pffl} \ee
where each $f_{\l_{iL}|\l_{iR}}$ has the regular presentation \eq{solnint}, the number operators have the specific realization
\begin{equation}
w_{1}=\frac{i}{8}\left( B_{\underline{\alpha \beta }}-P_{\underline{%
\alpha \beta }}\right) Y^{\underline{\alpha }}Y^{\underline{\beta }}\text{ \
, \ \ }w_{2}=\frac{i}{8}\left( B_{\underline{\alpha \beta }}+P_{%
\underline{\alpha \beta }}\right) Y^{\underline{\alpha }}Y^{\underline{\beta 
}}\text{ \ ,}  \label{w12B}
\end{equation}
with
\be B_{\underline{\alpha \beta }}=-(\G_{03})_{\underline{\alpha \beta }}\ ,\qquad P_{\underline{\alpha \beta }}=-(\G_{0'1})_{\underline{\alpha \beta }}\ ,\label{BPmatrices}\ee
and the creation/annihilation operators are  the linear combinations
\begin{gather}
a_{1}^+\ =\ \frac{1}{2}\left( y^{1}+\bar{y}^{\dot{1}}\right) \text{ , \ }%
a_{1}^-\ =\ \frac{i}{2}\left( y^{2}+\bar{y}^{\dot{2}}\right) \text{ ,}
\label{a1B} \\
a_{2}^+\ =\ \frac{i}{2}\left( y^{1}-\bar{y}^{\dot{1}}\right) \text{ , \ }%
a_{2}^-\ =\ \frac{1}{2}\left( y^{2}-\bar{y}^{\dot{2}}\right) \text{ .}
\label{a2B}
\end{gather}
Thus,
\be \Phi' \ = \ \sum_{\boldsymbol{\lambda }} \n_{\boldsymbol{\lambda }}f_{\boldsymbol{\lambda }}(Y)  \star \kappa_y  + {\rm conj} \ ,\label{Phitwist} \ee
where ${\rm conj}$ stands for the conjugate term required by reality conditions \eq{reality} (see \cite{ourBTZ} for the details), with the limitations \eq{lpmL}-\eq{lpmR} in the eigenvalues.

Now, fluctuation fields over the four-dimensional BTZ-like BGM background need to be left invariant by a full spatial transvection along the $S^1_K$ cycle. In the unfolded formalism this condition can be imposed on the fibre element $\Phi'$ (equivalently, $\Psi'$) \cite{ourBTZ} as 
\be \Phi ^{\prime } \ = \ \gamma ^{\prime -1}\star \Phi ^{\prime }\star \pi \left(
\gamma ^{\prime }\right) |_{\phi =2\pi}\text{ ,}  \ee
where $2\pi\sqrt{M}$ represents the circumference of the $S^1_K$ cycle of the BGM background, and 
\begin{equation}
\gamma ^{\prime } \ =  \ e_{\star }^{-\frac{i}{8}\sqrt{M}\phi P_{\underline{%
\alpha \beta }}Y^{\underline{\alpha }}Y^{\underline{\beta }}}=e_{\star }^{%
\frac{1}{2}\sqrt{M}\phi \left( w_{1}-w_{2}\right) }
\label{gamma is P}
\end{equation}
implements a finite transvection along the cycle. Imposing the identification condition on \eq{Phitwist} amounts to imposing it on each $f_{\boldsymbol{\l}}$, transforming as
\begin{eqnarray}
f_{\boldsymbol{\l}} &\longrightarrow &\gamma ^{\prime -1}\star f_{\boldsymbol{\l}}\star \gamma
^{\prime } = e^{\frac{1}{2}\sqrt{M}\varphi \left[ -\left( \lambda _{1L}-\lambda
_{2L}\right) +\left( \lambda _{1R}-\lambda _{2R}\right) \right] }f_{\boldsymbol{\l}} %
\text{ ,}
\end{eqnarray}
and requiring that the transformation is periodic in $%
\phi $ amounts to imposing the condition 
\begin{equation}
\left[ -\left( \lambda _{1L}-\lambda _{2L}\right) +\left( \lambda
_{1R}-\lambda _{2R}\right) \right] \in i\mathbb{R}\text{ .}
\end{equation}
Since we assume that $\lambda _{1,2\ R}+\frac{1}{2}\in \mathbb{Z}^{+}$,
this condition reduces to  
\begin{equation}
\mathrm{Re}\left( \lambda _{1L}-\lambda _{2L}\right) =\left( \lambda
_{1R}-\lambda _{2R}\right) \text{ .}\label{idreal}
\end{equation}
Furthermore, imposing that the transformation at $\phi =2\pi $ be the
identity, restricts
\begin{equation}
\mathrm{Im}\left[ \frac{\sqrt{M}}{2}\left( \lambda _{1L}-\lambda
_{2L}\right) \right] \in \mathbb{Z}\text{ .}\label{idim}
\end{equation}
Imposing also reality conditions and the bosonic projection $\pi\bar{\pi}(\Phi)=\Phi$ we finally reach the form of the Weyl 0-form integration constant that we shall employ,
\begin{eqnarray}
\Phi ^{\prime } 
   \ = \ &\displaystyle\sum_{\substack{ \text{All valid }  \\ \text{values of }%
\boldsymbol{\lambda }}}&\left[ \ \nu _{\boldsymbol{\lambda }}f_{\lambda _{1L}|\lambda _{1R}}\left(a_{1}^\pm\right) f_{\lambda _{2L}|\lambda _{2R}}\left( a_{2}^\pm \right)\right.\notag \\
%\star \kappa_{y}\right.\notag \\
%&&\ \ \ \ \ \ \ \ \ 
& & \displaystyle + \ \left. \left( \nu _{\boldsymbol{\lambda }}\right) ^{\ast }f_{-\lambda _{1R}|-\lambda _{1L}^{\ast }}\left(a_{1}^\pm\right)f_{-\lambda
_{2R}|-\lambda _{2L}^{\ast }}\left(  a_{2}^\pm\right)\right] \star \kappa _{y} \text{ ,} \label{phiprimefinal}
\end{eqnarray}
where 
\be \l_{iR}+\frac12\in\mathbb{Z}^+ \ ,\qquad  i=1,2\,; \label{constrlR} \ee
and both the real and the imaginary part of the left eigenvalues are quantized, and in particular 
\be {\rm Re}(\l_{iL})-\frac12\in \mathbb{Z} \ ,\qquad  {\rm with} \quad {\rm Re}(\l_{1L})-\l_{1R} \ = \ {\rm Re}(\l_{2L})-\l_{2R}\label{ident}\ee
and 
\be {\rm Im}(\l_{1L})\ = \ -{\rm Im}(\l_{2L})\in \frac{\mathbb{Z}} {\sqrt{M}}\ ,\label{imbos}\ee 
from which it follows that 
\be  \l_{1L}+\l_{2L}\ =  (\l_{1R}+\l_{2R}){\rm mod \,}2 \ . \label{rebos}\ee

We can now perform the star products \eq{LrotPhi} with the background gauge function in order to examine fluctuation fields in spacetime. Note that, as $AdS_4$ and (Ext)BGM are locally equivalent, in order to present the solution of the twisted-adjoint equation on a spacetime chart we can either $L_{AdS}$ or $L_{\rm (Ext)BGM}$, as the difference between the two will amount to a combined coordinate and local Lorentz transformation on the component fields. 

Either way, the final result reads, in $Sp(4;\R)$-covariant notation,
\be \Phi(x,Y)=L^{-1}\star \Phi ^{\prime }\star \pi \left( L\right)  =  \sum_{\substack{ \text{All valid }  \\ \text{values of }\boldsymbol{%
\lambda }}}\nu _{\boldsymbol{\lambda }}f_{\boldsymbol{\lambda }%
}^{L}\star \kappa _{y}+\text{ \ conj} \text{ ,}
\label{PhiL} \ee  
where $f^L_{\boldsymbol{\lambda }}=L^{-1}\star f_{\boldsymbol{\lambda }}\star L$, and
\bea  f^L_{\boldsymbol{\lambda }}\star \kappa_y &= & {\cal O}_{\boldsymbol{\lambda }_{1}}^{\varsigma_1}{\cal O}_{\boldsymbol{\lambda }_{2}}^{\varsigma_2}\int_{0}^{+\infty }d\tau _{1}\frac{\tau _{1}^{\lambda _{1R}-\lambda _{1L}-1}%
}{\Gamma \left( \lambda _{1R}-\lambda _{1L}\right) }\int_{0}^{+\infty }d\tau _{2}\frac{\tau _{2}^{\lambda
_{2R}-\lambda _{2L}-1}}{\Gamma \left( \lambda _{2R}-\lambda _{2L}\right) } \nn\\
&&\times \ \  \frac{1}{\sqrt{\det \check \vark^L}}\, \exp\left[-\frac1{2}(\ty^L-i\theta^L)\frac{\check \vark^L}{\det \check \vark^L}(\ty^L-i\theta^L)+\frac{1}2\yb\check \varkb^L\yb-\bar \theta^L\yb\right] \ , \label{Phicov}\eea
in which:

\begin{itemize}

\item we have introduced the shorthand notation 
\be {\cal O}_{\boldsymbol{\lambda }_{i}}^{\varsigma_i} \ := \ \oint_{C(\pm1)}\frac{d\varsigma _{i}}{2\pi i}\frac{\left( \varsigma
_{i}+1\right) ^{\lambda _{iL}-\frac{1}{2}}}{\left(\varsigma _{i}-1\right)
^{\lambda _{iR}+\frac{1}{2}}}  \ ; \label{short}\ee

\item we define the modified oscillators $\ty^L:=y-i\check v^L\yb$;

\item the spacetime-dependent matrices $(v^L)_{\a\bd}$,  $(\vark^L)_{\a\b}$ and $(\varkb^L)_{\ad\bd}$ are the $2\times 2$ blocks of the matrix %
\bea
\check K^L(\varsigma_1,\varsigma_2; Y) \ :=  \ \frac{\varsigma_1+\varsigma_2}2 K^L_{(+)}+\frac{\varsigma_2-\varsigma_1}2  K^L_{(-)} % \notag\\
\ = \   \ -\frac18\left[ y\check\varkappa^L y + \yb\check{\bar \varkappa}^L\yb+2 y \check v^L \yb  \right] \ ,
\eea 
where
\bea &\displaystyle K^L_{(q)} \ = \ -\frac18 Y^{L\underline{\a}}K_{(q)\underline{\a}}{}^{\underline{\b}}Y^L_{\underline{\b}} \ = \ -\frac18 Y^{\underline{\a}}K^L_{(q)\underline{\a}}{}^{\underline{\b}}Y_{\underline{\b}} \ ,& \nn\\
&\displaystyle K^L_{(q)\underline{\a}}{}^{\underline{\b}} \ = \ -\left(L^T K_{(q)} L\right)_{\underline{\a}}{}^{\underline{\b}} \ = \ \left(\begin{array}{cc}\vark^L_{(q)\a\b} & v^L_{(q)\a\bd} \\ 
\bar v^L_{(q)\ad\b} & \bar \vark^L_{(q)\ad\bd}\end{array}\right) \ ,& \\ &K_{(+)}=iB \ , \qquad \ K_{(-)}=iP \label{KBP}&\eea
follow from the $L$-rotation of the rigid matrices \eq{BPmatrices} appearing in $\Phi'$ in the linear combinations \eq{w12B}, the matrix $L$ enters via \eq{YLrotn}, and $(\theta^{L\a},\bar{\theta}^{L\ad})$ are linear-in-$\t_i$ and $x$-dependent spinors. 
The precise expressions for all these quantities can be found in \cite{ourBTZ}. 

\end{itemize}

We shall soon specify this general expression of the Weyl zero-form to a concrete case, but one important remark that we can make at this stage is that the star products with the gauge function render $\Phi(x,Y)$ a \emph{regular} function of $Y$ at generic spacetime points. This is non-trivial, considering that, in order to  have non-trivial momentum on $S^1_K$ we had to allow for complex powers of the oscillators, and that the latter lead, in the integral presentation of $\Phi'$ \eq{phiprimefinal} with \eq{solnint}, to  ill-defined $\tau$-integrals for $Y=0$. However, displacing the Weyl zero-form away from the unfolding point, by means of the star products with $L$, leads to the appearance of a terms bilinear in $\theta^L$, i.e. bilinear in $\t_i$, at the exponent of the integrand in \eq{Phicov}: this helps the convergence of the Mellin transforms and restores analyticity at $Y=0$ (at least for generic spacetime points), which means that \eq{Phicov} can be considered a proper generating function of fluctuation fields according to \eq{Phicomp} \cite{ourBTZ}. 

Let us now extract the $s=0$ component, viz.
\begin{equation}
C(x) \ := \ \ f^L_{\boldsymbol{\lambda }}\star \kappa_y|_{Y=0} + {\rm c.c.}\ , \label{scal}
\end{equation} 
and compare with the result \eq{solmetric} obtained in metric-like formalism. In order to do so, we must choose eigenvalues $\boldmath{\l}$ such that $\Phi'$ has vanishing eigenvalue under $iB$ and eigenvalue $in/\sqrt{M}$ under $iP$, according to \eq{twadjeig} (with the identifications \eq{KBP} and $n \to \l$). The simplest such choice is 
\begin{equation}
\lambda _{1L}=\frac{1}{2}+i\,\frac{n}{\sqrt{M}}\ \ ,\ \ \ \lambda _{2L}=\frac{1%
}{2}-i\,\frac{n}{\sqrt{M}}\ \ ,\ \ \ \lambda _{1R}=\lambda _{2R}=\frac{1}{2}\
\ ,\ \ \ (n\in \mathbb{Z)}\text{ ,}
\label{eigenchoice}
\end{equation}
which is compatible with the constraints \eq{constrlR}-\eq{rebos}. With this choice, defining $p:=\frac{n}{\sqrt{M}}$, the scalar field takes the form
\bea C(x) = \  \int_{0}^{+\infty }d\tau _{1}\frac{\tau _{1}^{-ip-1}}{\Gamma \left(
-ip\right) }\ \int_{0}^{+\infty }d\tau _{2}\frac{\tau _{2}^{ip-1}}{\Gamma
\left( ip\right) } \frac{1}{\sqrt{\D^2}}\, e^{-\frac1{2\D^2 }(a\t^2_1+b\t_1\t_2+c\t^2_2)} +{\rm c.c.} \label{scalC} \ ,
\eea
where $a,b,c$ are coefficients depending on spacetime coordinates (see \cite{ourBTZ}) and ${\rm c.c.}$ is the complex conjugate, required by the reality conditions \eq{reality}. Computing the two remaining integrals finally gives, in the same coordinates used for \eq{solmetric},
\be C 
=  
 e^{-in\phi}\frac{\cosh \left\{ \frac{n}{\sqrt{M}}\arcsin %
\left[ \sqrt{\frac{M\D^2}{\xi^2}}\right] \right\} } {\sqrt{\D^2}}+{\rm c.c.}\ .\label{C1C2}\ee
Indeed, recalling that $\D^2=\xi^2/M-x^2$, and using the identity ${\rm arctanh \frac{x}{\sqrt{-\D^2}}}= -i\,{\rm arcsin}\sqrt{\frac{M\D^2}{\xi^2}}$, we can see that, apart from having here subjected $C$ to be real in accordance with the reality conditions on $\Phi$, the solution here constructed coincides with \eq{solmetric} with $c_2=0=c_4$, $c_1c_3=i$.

As previously found in the metric-like formalism, the scalar field has a membrane-like singularity on the surface $\Delta^2= 0$. Moreover, approaching the singularity of the (Ext)BGM background, \emph{i.e.} in the limit $\xi\to 0$, the scalar field remains bounded but becomes indefinite, as it oscillates with diverging frequency. However, the singularities need to be re-examined at the level of the master field $\Phi(x,Y)$. Taking into account that, as \eq{short}-\eq{KBP} exhibit, for our choice of eigenvalues  $\det\check\vark^L=\D^2$, we can see from \eq{PhiL}-\eq{Phicov} that for $\D^2\to 0$ the integrand behaves as a delta-sequence, and indeed it is possible to show that 
\be  \lim_{\D^2\to 0} \Phi \ \propto \  f(X) {\cal O}_{\boldsymbol{\l}_1} {\cal O}_{\boldsymbol{\l}_2} \d^2(\hat y) \ .  \ee
where $f(X)$ is a function of the spacetime coordinates and $\hat y:=\lim_{\D^2\to 0}\ty^L$ are non-commuting oscillators (see Appendix D in \cite{ourBTZ}). This means that, much like for the Weyl singularity of the Schwarzschild higher-spin black hole, the membrane-like singularities of individual individual spin-$s$ Weyl tensors coalesce into a delta-function behaviour of the corresponding master field on that surface. Therefore, as remarked above, the fluctuation Weyl zero-form master field remains well-defined as a star-product algebra element, and in this sense experience the membrane singularity as a smooth surface. 

Furthermore, one can observe that $\D^2 |_{x=0}=\xi^2/M$. 
The analysis of the membrane-like singularity therefore suggests that also $\xi=0$ is a regular point, in the sense that the master field is given here by a well-defined regular prescription.
For these reasons, recalling that the unfolded field equations never require to invert the Vielbein, we expect that the master field configuration and the differential algebra that defines its dynamics in the unfolded approach can be continued through the causal singularity of the BGM manifold, thus exploring the full background manifold ${\rm ExtBGM}=AdS_3\times_\xi S^1_K$. 
We leave the full exploration of this issue for future research.

\acknowledgments

\noindent It is a pleasure to thank the Organizers of CORFU2021, in particular Harold Steinacker and George Zoupanos, for arranging a stimulating and very pleasant workshop.  
A large part of the results reported on in this paper have been obtained in collaboration with R. Aros, D. De Filippi and Y. Yin, whom we would like to thank. We are also grateful to M. Bianchi, F. Diaz, G. Pradisi, B. Vallilo for stimulating discussions.

\appendix

\scs{AdS and spinor conventions}\label{app:conv}

We use the conventions of \cite{fibre} in which $SO(2,3)$ generators $M_{AB}$ with $A,B=0,1,2,3,0'$ obey
\be [M_{AB},M_{CD}]\ =\ 4i\y_{[C|[B}M_{A]|D]}\ ,\qquad
(M_{AB})^\dagger\ =\ M_{AB}\ ,\label{sogena}\ee
which can be decomposed using $\eta_{AB}~=~(\eta_{ab};-1)$ with $a,b=0,1,2,3$ as
\be [M_{ab},M_{cd}]_\star\ =\ 4i\y_{[c|[b}M_{a]|d]}\ ,\qquad
[M_{ab},P_c]_\star\ =\ 2i\y_{c[b}P_{a]}\ ,\qquad [P_a,P_b]_\star\ =\
i\lambda^2 M_{ab}\ ,\label{sogenb}\ee
where $M_{ab}$ generate the Lorentz subalgebra $\mso(1,3)$, and $P_a=\l M_{0'a}$ with $\l$ being the inverse $AdS_4$ radius related to the cosmological constant via $\L=-3\l^2$. We set $\l=1$ in the following, as we do in the body of the paper.

Decomposing further with respect to the maximal compact subalgebra $\mso(2)\oplus\mso(3)$, generated by the compact $AdS_4$ energy generator $E=P_0=\l M_{0'0}$ and the spatial rotation generators $M_{rs}$ with $r,s=1,2,3$, the remaining generators then arrange into energy-raising and energy-lowering combinations identified with
\bea L^\pm_r&=& M_{0r}\mp iM_{0'r}\ =\ M_{0r}\mp iP_r\
,\label{Lplusminus}\eea
leading to the following $E$-graded decomposition of the
commutation rules \eq{sogena}:
\bea & [L^-_r,L^+_s] \ = \ 2iM_{rs}+2\d_{rs}E \ , \ ,\  \qquad \
[M_{rs},M_{tu}] \ = \ 4i\d_{[t|[s}M_{r]|u]} \ ,& \label{mm}\\[5pt]
& [E,L^{\pm}_r] \ = \ \pm L^{\pm}_r\ ,\  \qquad \ [M_{rs},L^\pm_t]\ = \
2i\d_{t[s}L^\pm_{r]}\ .&\label{ml}\eea
The generators $(E,M_{rs},L^\pm_r)$ are also referred to as generators of the \emph{compact basis}, or \emph{compact split} of $\mso(2,3)$.

In terms of the oscillators $Y_{\underline\a}=(y_\a,\yb_{\ad})$, the realization of the generators of $\mso(2,3)$ is taken to be
\be M_{AB}~=~ -\ft18  (\G_{AB})_{\underline{\a\b}}\,Y^{\underline\a}\star Y^{\underline\b}\ ,\label{MAB}\ee
 \be
 M_{ab}\ =\ -\frac18 \left[~ (\s_{ab})^{\a\b}y_\a\star y_\b+
 (\bar{\sigma}_{ab})^{\ad\bd}\bar y_{\ad}\star \yb_{\bd}~\right]\ ,\qquad P_{a}\ =\
 \frac{1}4 (\s_a)^{\a\bd}y_\a \star \yb_{\bd}\ ,\label{mab}
 \ee
using Dirac matrices obeying $(\Gamma_A)_{\underline\a}{}^{\underline\b}(\Gamma_B)_{\underline{\b\gamma}}=
\eta_{AB}C_{\underline{\a\gamma}}+(\Gamma_{AB} C)_{\underline{\a\gamma}}$, 
\begin{equation}
\left( \Gamma_{0'a}\right) _{\underline{\alpha }}^{\ \ \underline{\beta }
}\equiv \left( \Gamma_{a}\right) _{\underline{\alpha }}^{\ \ \underline{\beta }
}=\left( 
\begin{array}{cc}
0 & -\left( \sigma_{a}\right) _{\alpha }^{\ \ \dot{\beta}} \\ 
-\left( \bar{\sigma}_{a}\right) _{\dot{\alpha}}^{\ \ \beta } & 0%
\end{array}
\right) \text{ ,}
\end{equation}
and
\begin{equation}
\left( \Gamma _{ab}\right) _{\underline{\alpha \beta }}=\left( 
\begin{array}{cc}
\left( \sigma _{ab}\right) _{\alpha \beta } & 0 \\ 
0 & \left( \bar{\sigma}_{ab}\right) _{\dot{\alpha}\dot{\beta}}%
\end{array}
\right) \text{ .}
\end{equation}
and van der Waerden symbols obeying
 \be
  (\s^{a})_{\a}{}^{\ad}(\bar{\sigma}^{b})_{\ad}{}^{\b}~=~ \y^{ab}\d_{\a}^{\b}\
 +\ (\s^{ab})_{\a}{}^{\b} \ ,\qquad
 (\bar{\sigma}^{a})_{\ad}{}^{\a}(\s^{b})_{\a}{}^{\bd}~=~\y^{ab}\d^{\bd}_{\ad}\
 +\ (\bar{\sigma}^{ab})_{\ad}{}^{\bd} \ ,\label{so4a}\ee\be
 \ft12 \e_{abcd}(\s^{cd})_{\a\b}~=~ i (\s_{ab})_{\a\b}\ ,\qquad \ft12
 \e_{abcd}(\bar{\sigma}^{cd})_{\ad\bd}~=~ -i (\bar{\sigma}_{ab})_{\ad\bd}\ ,\label{so4b}
\ee
\be ((\s^a)_{\a\bd})^\dagger~=~
(\bar{\sigma}^a)_{\ad\b} ~=~ (\s^a)_{\b\ad} \ , \qquad ((\s^{ab})_{\a\b})^\dagger\ =\ (\bar{\sigma}^{ab})_{\ad\bd} \ .\ee
and raising and lowering spinor indices according to the
conventions $A^\a=\epsilon^{\a\b}A_\b$ and $A_\a=A^\b\epsilon_{\b\a}$ where
\be \e^{\a\b}\e_{\g\d} \ = \ 2 \d^{\a\b}_{\g\d} \ , \qquad
\e^{\a\b}\e_{\a\g} \ = \ \d^\b_\g \ ,\qquad (\e_{\a\b})^\dagger \ = \ \e_{\ad\bd} \ .\ee
In order to avoid cluttering the expression with many spinor indices, in the paper we also use the matrix notations 
\bea & A^{\underline{\a}} B_{\underline{\a}} \ = :\  AB \ = \ ab+\bar a \bar b \ := \ a^\a b_\a + \bar a^{\ad}\bar b_{\ad} \ ,&\\
& aMb \ := \ a^\a M_{\a}{}^\b b_\b \ , \qquad aN\bar{b} \ := \ a^\a N_{\a}{}^{\bd} \bar{b}_{\bd} \ . &
\eea
The $\mso(2,3)$-valued connection
 \be
  \O~:=~-i \left(\frac12 \omega^{ab} M_{ab}+e^a P_a\right) ~:=~ \frac1{2i}
 \left(\frac12 \omega^{\a\b}~y_\a \star y_\b
 +  e^{\a\dot\b}~y_\a \star {\bar y}_{\dot\b}+\frac12 \bar{\omega}^{\dot\a\dot\b}~{\bar y}_{\dot\a}\star {\bar y}_{\dot\b}\right)\
 ,\label{Omega}
 \ee
  \be
 \o^{\a\b}~=~ -\ft14(\s_{ab})^{\a\b}~\o^{ab}\ , \qquad \omega_{ab}~=~\ft12\left( (\s_{ab})^{\a\b} \o_{\a\b}+(\bar\s_{ab})^{\ad\bd} \bar\o_{\ad\bd}\right)\ ,\ee
 \be e^{\a\dot\a}~=~ \ft{1}2(\s_{a})^{\a \dot\a}~e^{a}\ , \qquad e_a~=~ - (\s_a)^{\a\ad} e_{\a\ad}\ ,\label{convert}\ee
and field strength
\bea {\cal R} & := & d\O+\O\star \O~:=~-i \left(\frac12 {\cal R}^{ab}M_{ab}+{\cal R}^a P_a\right) \nn\\[5pt]
& :=&  \frac1{2i}
 \left(\frac12 {\cal R}^{\a\b}~y_\a \star y_\b
 +  {\cal R}^{\a\dot\b}~y_\a \star {\bar y}_{\dot\b}+\frac12 \bar{\cal R}^{\dot\a\dot\b}~{\bar y}_{\dot\a}\star {\bar y}_{\dot\b}\right)\
 ,\label{calRdef}\eea
\be
 {\cal R}^{\a\b}\ =\ -\ft14(\s_{ab})^{\a\b}~{\cal R}^{ab}\ ,
 \qquad {\cal R}_{ab}~=~\ft12\left( (\s_{ab})^{\a\b} {\cal R}_{\a\b}+(\bar\s_{ab})^{\ad\bd} \bar{\cal R}_{\ad\bd}\right)\ ,\ee
 \be
 {\cal R}^{\a\dot\a}\ =\ \ft{1}2(\s_{a})^{\a \dot\a}~{\cal R}^{a}\ ,
 \qquad {\cal R}_a~=~ - (\s_a)^{\a\ad} {\cal R}_{\a\ad}\ .\ee
In these conventions, it follows that
 \be
 {\cal R}_{\a\b}~=~ d\o_{\a\b} -\o_{\a}^{\g}\o_{\g\b}-
 e_{\a}^{\dot{\g}}\bar e_{\dot{\g}\b}\ ,\qquad
 {\cal R}_{\a\dot\b}~=~  de_{\a\bd}+ \o_{\a\g}\wedge
 e^{\g}{}_{\bd}+\bar{\o}_{\bd\dd}\wedge e_{\a}{}^{\dd}\
 ,\ee\be
 {\cal R}^{ab}~=~ R_{ab}+
 e^a\wedge e^b\ ,\qquad R_{ab}~:=~d\o^{ab}+\o^a{}_c\wedge\o^{cb}\ ,\ee\be
 {\cal R}^a~=~ T^a ~:=~d e^a+\o^a{}_b\wedge e^b\ ,
 \label{curvcomp} \ee
where $R_{ab}:=\frac12 e^c e^d R_{cd,ab}$ and $T^a:=e^b e^c T^a_{bc}$ are the Riemann and torsion two-forms.

\scs{$U(1)$ subgroups and Klein operators}\label{App:proof}

A symmetric $2n\times 2n$ matrix $R$ that is a square root of the identity, in the sense that $R^2=1$, induces a split of the $2n$-dimensional symplectic coordinates $Y$ into a pair of $n$-dimensional canonical coordinates $(Y_+,Y_-)$ by means of projectors $P_\pm$ such that
\be
  \label{eq:XpmR}
  Y_{\pm}
  :=
  P_{\pm}Y\ ,\qquad P_{\pm }
  :=
  \frac{1}{2}\left(1\pm R\right)\ , \qquad R Y_\pm=\pm Y_\pm \ ,
\ee
satisfying
\be  [Y_{\epsilon }^{I},Y_{\epsilon' }^{J}]_\star= 2i\epsilon'  \delta_{\epsilon,-\epsilon'} P_{\epsilon }^{IJ}\ ,\qquad \epsilon,\epsilon'=\pm
\ .\ee
Then
\be w_Y := \frac{i}{4}  Y R Y\ee
is the Weyl-ordered number operator,
\be [ w_Y, Y_\pm]_\star = \pm  Y_\pm \ ,\ee
differing from the corresponding normal-ordered counterpart ${\cal N}_Y=-\frac{i}2 Y_+\star Y_-$ by an ordering constant,
\be w_Y={\cal N}_Y+\frac{n}2  \ .\label{wN}\ee

As is well-known \cite{Folland,Guillemin:1990ew,Woit:2017vqo} the operator $\exp_\star(i\theta w_Y)$ generates a $U(1)$ subgroup of the metaplectic group $Mp(2n,\R)$, and, for $\theta=2\pi$, $\exp_\star(2\pi i w_Y)=(-1)^n\exp_\star(2\pi i{\cal N}_Y)$, corresponding to the characteristic sign of the metaplectic representation in its action on a Fock space\footnote{Indeed, the exponential of normal ordered generators is a true, not two-valued, representation of $U(1)$, and in fact of the whole $U(n)$ included in $Sp(2n,\R)$, which however does not extend to the entire $Sp(2n,\R)$ as it is the case instead for the two-valued metaplectic representation.}. Note that in the four-dimensional case treated in this paper (i.e., $n=2$) the split induced by such an $R$ gives rise to a $2D$ Fock space, in which the action of $\exp_\star(2\pi i w_Y)$ is $2\pi$-periodic.

Due to the defining properties of the gamma matrices (collected in Appendix \ref{app:conv}), the combination $n^a\G_a$ with $n^a n_a=1$, that appeared in Section \ref{sec:vacua}, is one such $R$-matrix. In particular, in view of the realization \eq{mab},
\be \exp_\star (\a n^a P_a) \ = \ \exp_\star\left(-\frac{i\a}{2}\,w_Y\right) \ , \ee
which, upon identifying the full $Y$-dependent inner Klein operator $ K_Y=\exp_\star(\pm i\pi {\cal N}_Y)=\kappa_y\star\kb_{\yb}$ (see \cite{meta} for the details of the identification) and using \eq{wN} in the case $n=2$, explains Eq. \eq{T-2pi}. 

A special case is the operator $\exp_\star(2\pi i E)$, appearing in Eq. \eq{holkappa}:
\be \exp_\star(2\pi i E) \ = \ \exp_\star( -i\pi w_Y) \ = \  -\exp_\star( -i\pi  {\cal N}_Y) \ = \ - \kappa_y\star\kb_{\yb} \ . \ee

\end{document}